\documentclass[a4paper,preprint,aps,superscriptaddress]{revtex4}

\usepackage{graphicx}
\usepackage{dcolumn}
\usepackage{bm}

\newcommand{\eqa}{\begin{equation}}
\newcommand{\eqz}{\end{equation}}
\newcommand{\eqma}{\begin{eqnarray}}
\newcommand{\eqmz}{\end{eqnarray}}

\begin{document}
\title{From {\it ab initio} quantum chemistry to molecular dynamics:\\
The delicate case of hydrogen bonding in ammonia}
\author{A. Daniel Boese}
\affiliation{Department of Organic Chemistry,
Weizmann Institute of Science, 76100 Re\d{h}ovot, Israel}
\author{Amalendu Chandra$^{*}$}
\affiliation{Lehrstuhl f\"ur Theoretische Chemie, Ruhr-Universit\"at Bochum,
44780 Bochum, Germany}
\author{Jan M. L. Martin}
\affiliation{Department of Organic Chemistry,
Weizmann Institute of Science, 76100 Re\d{h}ovot, Israel}
\author{Dominik Marx$^2$}
\date{305335JCP: Received May 15, 2003; Accepted June 18, 2003}
\smallskip
\begin{abstract}
\indent
The ammonia dimer (NH$_3$)$_2$ has been investigated using
high--level {\it ab initio} quantum chemistry 
methods and density functional theory
 (DFT). 
%
%
The structure and energetics of important isomers is obtained
to unprecedented accuracy without resorting to experiment.  
The global minimum of eclipsed $C_s$ symmetry is characterized
by a significantly bent hydrogen bond which deviates from linearity by 
as much as $\approx 20^\circ$.   
In addition, the so--called cyclic $C_{2h}$ structure, resulting
from further bending which leads to two equivalent ``hydrogen bonding contacts'', 
is extremely close in energy on an overall flat potential energy surface.  
It is demonstrated that none of the currently available
(GGA, meta--GGA, and hybrid) 
density functionals satisfactorily describe
the structure and relative energies of this nonlinear hydrogen bond.
We present a novel density functional, HCTH/407+, which is
designed to describe this
sort of hydrogen bond quantitatively on the level of the dimer,
contrary to e.g. the widely used BLYP functional. 
This improved
generalised gradient approximation (GGA) 
functional is employed in Car--Parrinello {\it ab initio} molecular
dynamics simulations of liquid ammonia to judge its performance in describing the 
associated liquid. 
%
Both the HCTH/407+ and BLYP functionals describe the
properties of the liquid well as judged by
analysis of
radial distribution functions, hydrogen bonding structure and dynamics,
translational diffusion, and orientational relaxation processes.
It is demonstrated that the solvation shell of the ammonia molecule 
in the liquid phase is dominated 
by steric packing effects and not so much by directional 
hydrogen bonding interactions. 
In addition, 
the propensity of ammonia molecules to form bifurcated and multifurcated 
hydrogen bonds in the liquid phase is found to be negligibly small. 
%

\vspace{3cm}
-----------------------------\\
$^{*}$ On 
sabbatical 
leave from: Department of Chemistry, Indian Institute of Technology, Kanpur, India 208016.
\end{abstract}
\maketitle  

\section{Introduction and Motivation}


Wavefunction--based quantum chemistry methods 
as well as density functional theory (DFT) 
are now well established electronic structure 
techniques within Theoretical Chemistry~\cite{TC}.
The former is traditionally used to investigate small finite systems,
i.e. ``gas phase problems'',
whereas the latter had its first successes in solid state physics,
i.e., in the framework of ``periodic lattice problems''. 
%
%
There is also ample evidence that
wavefunction--based quantum chemistry methods are able to 
predict properties of molecular systems up to virtually 
any desired accuracy as a result of systematically improving 
the correlation treatment together with the wavefunction representation.
However, the numerical complexity, as e.g. measured by the computer time 
needed for such studies, explodes easily {\em in praxi} either before 
reaching a sufficient level of accuracy or before treating
systems that are sufficiently large.  
DFT methods, on the other hand, are known to lead
to astonishingly good predictions for a wide variety of 
systems at moderate cost~-- of course dramatic failures are also well documented
in the literature.
When describing molecular solids, associated liquids, or large biosystems,
noncovalent interactions such as hydrogen bonds and van der Waals
interactions might play an important role.
%
The ideal approach would be to use post Hartree-Fock methods
converged to the basis set limit,
which turns out to be computationally
prohibitive, especially when used in conjunction with molecular dynamics. 
While simple DFT methods are unable to correctly
describe van der Waals interactions
\cite{Rappe,Gdanitz,Hirao},
hydrogen bonds are known to be well within the capabilities of
DFT approaches once suitable gradient--corrected functionals
are devised.
A pioneering step in the early nineties was the investigation of 
hydrogen bonds in the water dimer~\cite{kari1,sim92,cpwatern}
and in small water clusters~\cite{kari2}
%
based on such functionals, 
which are of the generalised gradient approximation (GGA) type.
Soon thereafter it was demonstrated that also the structure 
and dynamics of {\em liquid bulk} water 
can be described faithfully~\cite{cpwatern}.
Now there is ample evidence that various properties of water 
can be computed quite reliably
using GGA--type functionals, see e.g. 
Refs.~\cite{cpwatern,tuck,Sprik,SP,SBP,MTHP,prr00,pbe-water,pbe-water2,%
HCTH120,MTP,krack,IV} 
and Refs.~\cite{Pais,TBH,Boyd,tHCTH} for dimer calculations.  
But how about other hydrogen bonded systems?

Among hydrogen--bonded dimers the ammonia dimer is peculiar since it features a rather
unusual hydrogen bond, see Refs.~\cite{nelson87,havenith} for an early and a
very
recent overview.
Instead of an essentially linear hydrogen bond, similar to the one of the water dimer
having a H--O$\cdots$O angle of only 5.5$^\circ$~\cite{klopper00}, the hydrogen bond
%
%
in the ammonia dimer is significantly bent as depicted in Fig.~\ref{fig1}c or d.
Furthermore, close in energy is the so--called cyclic structure where an even more pronounced
bending of the hydrogen bond produces two equivalent ``hydrogen bonding contacts''
according to Fig.~\ref{fig1}e.
As an interlude we stress that such strongly bent hydrogen bonds play an important role in
many biological systems, in particular if the bending allows for more than one interaction of
a donor with acceptor molecules which leads to so--called bifurcated or multifurcated
hydrogen bonds~\cite{steiner02,noteHB}.
In carbohydrates over 25\% of the hydrogen bonds turn out to be multifurcated~\cite{steiner02}
and in proteins~\cite{Saenger}  about 40 and 90\% of the hydrogen bonds in $\beta$--sheet and
$\alpha$--helix structures, respectively, are of this type.
In this context the ammonia dimer could serve as a prototype system in order to investigate
nonlinear hydrogen bonding on the level of a minimal model, i.e. a free dimer in the gas phase.

Guided by analogy or chemical intuition, one might guess that the ammonia dimer possesses a
``classical'' quasi--linear hydrogen bond similar to other dimers such as those of
water or hydrogen fluoride.
Early calculations, see e.g. Refs.~\cite{lin1,rlin1,rlin2} and references cited therein,
actually assumed from the outset a perfectly linear bond similar to the one shown in  Figs.
\ref{fig1}a or b or predicted a quasi--linear hydrogen bond as depicted in Figs.~\ref{fig1}c or d, 
while others~\cite{C2h,ahlrichs,cyclic1} favoured strongly bent up to cyclic dimers similar
to Figs.~\ref{fig1}c--e. This controversy got fueled by microwave measurements~\cite{MW}
providing evidence for a 
quasi--rigid 
cyclic structure such as the one in Fig.~\ref{fig1}e.
The issue of a quasi--linear hydrogen bond vs. a cyclic structure remained controversial
(see e.g. Refs.~\cite{KT,rlin3}) 
up to early nineties where both experiment and theory started to converge. 
Thanks to sophisticated 
experiments~\cite{hav91,hav92,loeser92,say-vrtrev,DP,heineking95,linnartz95,cotti96} 
and extensive
computations~\cite{Olthof,Olthof2} there is now consensus~\cite{havenith} 
that the ammonia dimer is a {\em fluxional} molecular complex
with an {\em equilibrium structure} that 
is characterized by a bent hydrogen bond 
such as the one shown in Fig.~\ref{fig1}c.
Combining far--infrared spectra with static and dynamic calculations~\cite{Olthof,Olthof2,noteFIX}, 
the H--N$\cdots$N hydrogen bond angle (between the N--H bond of the proton donor and the N$\cdots$N axis) 
was determined to be 26$^\circ$
in the equilibrium structure with a N$\cdots$N distance of $\approx 3.354$~{\AA}.
%
Note, however, that this ``experimental'' structure of the ammonia dimer was deduced within the
rigid monomer approximation~\cite{Olthof,noteFIX}, which might introduce a systematic bias.
In fact, a recent
study~\cite{saykally02} of the water dimer has shown that monomer flexibility is indeed important
in determining its vibration--rotation--tunneling spectrum and thus
``experimental'' structures derived from it.
It is expected that such monomer flexibility would also be important for
the ammonia dimer.
Furthermore, 
the global potential energy surface is found to be very flat~\cite{Olthof,Olthof2}, 
e.g. the cyclic C$_{2h}$ structure
was estimated to be about only 7~cm$^{-1}$ higher in energy, which is qualitatively consistent with the observation 
that the dimer is a very fluxional complex.
Most interestingly, there is now experimental
support~\cite{nh3super} advocating
a more cyclic equilibrium structure for (NH$_3$)$_2$ in superfluid Helium droplets at 0.4~K.

More recent wavefunction--based quantum chemical calculations~\cite{LeePark,Lindh} 
on the structure and energetics of the
ammonia dimer support in some aspects the earlier conclusions~\cite{Olthof,Olthof2,havenith}. 
Second--order M{\o}ller--Plesset perturbation calculations (MP2) with medium--sized basis sets
(for geometry optimization followed by MP4 single--point energy calculations)\cite{LeePark}
resulted however in a significantly smaller H--N$\cdots$N hydrogen bond angle of 12$^\circ$ together with a
reasonable low barrier of 7.6~cm$^{-1}$. Most recently, another (valence--only) MP2 study~\cite{Lindh},
this time
near the one--particle basis set limit, was reported.
At the MP2 limit, the hydrogen bond angle of 23$^\circ$ was closer to the 
recommended value~\cite{Olthof,Olthof2}, but the barrier of 24~cm$^{-1}$ seems too high and 
the dissociation energy $D_{\rm e}$ (into separated and optimized monomer fragments) was estimated to
be $13.5 \pm 0..3 $~kJ/mol.
This clearly hints to the importance of higher--order correlation effects and core--valence corrections,
in particular since they account for 0.4~kJ/mol in the case of the water dimer~\cite{Klopper2}
(which already exceeds the error bar given in Ref.~\cite{Lindh}).
DFT calculations of the ammonia dimer are both rather scarce~\cite{Zhu,Suhai,Jursic,DMT}
and somewhat inconclusive as to what level of accuracy DFT methods describe this intermolecular
interaction.
In the following we will clearly demonstrate that none of the 14 
widely used GGA,
meta--GGA, and hybrid functionals checked by us lead to a satisfactory description of both the 
structure and relative energies of the ammonia dimer 
(despite large basis sets being used and basis set superposition error being corrected for).
%
Thus, even at the level of the ammonia dimer there is still
room for improvement based on purely theoretical grounds~!

Thus far we considered the behavior of ammonia in the limit of forming a dimer,
e.g. in the very dilute gas phase.
The question arises, if or how its peculiar behavior concerning hydrogen--bonding
in the dimer is present in its condensed phases?
Indeed, based on the {\em apparent} stability 
and rigidity
of the {\em cyclic} dimer in the gas phase~\cite{MW} 
it was concluded in quite general terms  
some time ago 
that NH$_3$ might well be best described 
as a powerful hydrogen--bond acceptor with little propensity 
to donate hydrogen bonds~\cite{nelson87};
note that now it is however recognized that the 
ammonia dimer is non--rigid and non--cyclic~\cite{havenith}. 
The x--ray structure of {\em crystalline} ammonia at low temperatures~\cite{CS1,CS2} features a
staggered, directed hydrogen bonded network, with the monomers being arranged
like the dimer in Fig.~\ref{fig1}d.  
Furthermore, based on neutron powder diffraction experiments, it was proposed that solid 
ammonia has an ``unusual shared--hydrogen bond geometry''~\cite{nelmes96} and, in
particular,  the high--pressure phase ND$_3$--{\em IV} 
contains bifurcated hydrogen bonds with bending angles as large as about 30$^\circ$. 
A most recent theoretical study~\cite{fortes03}, however, did not support the presence of
bifurcated hydrogen bonds in phase {\em IV} of solid ammonia. 
In this latter study, it was concluded 
through calculations of electron density between intermolecular nitrogen and hydrogen atoms 
that the apparent bifurcated hydrogen bond geometry in solid ammonia {\em IV} is actually 
a single hydrogen bond perturbed by a neighboring interacting atom. 
The structure of {\em liquid} ammonia
has been studied by x--ray~\cite{AHN} and neutron~\cite{CB} diffraction for a long time. 
Only the total structure factor and the total radial distribution function
could be extracted in these early experiments and the presence of hydrogen bonds
in liquid ammonia was inferred from a comparison of the experimental structure
factor with the known structure of the solid.
More recently, 
a set of neutron diffraction experiments was performed~\cite{RNR} 
on liquid NH$_{3}$, ND$_{3}$, and an isomolar NH$_{3}$/ND$_{3}$ mixture. 
The powerful isotopic substitution technique, in conjunction
with sophisticated data analysis, allowed
extraction of
all partial radial distribution functions at 
213 and 273~K. 
Contrary to earlier
conceptions
(as
presented in many standard textbooks on that subject) 
the spatial arrangement of nitrogen atoms 
showed that no extended hydrogen bonded network exists in liquid ammonia. 
Nevertheless, some 
degree of hydrogen bonding was inferred from the 
temperature dependence of 
the N-H and H-H radial distribution functions.  
However, the hydrogen bond interaction in 
liquid ammonia proved to be much weaker than that in water
and no clear hydrogen bond peak
was observed in either N-H or H-H correlations, unlike the case of water.

Due to the importance of liquid ammonia and in particular 
its metallic solutions, 
also many theoretical attempts have been made in past two decades to
understand its liquid phase.
Most of these 
studies are based on classical Monte Carlo (MC) or molecular dynamics (MD) 
simulations using 
empirical site--site \cite{WLJ,KS1,IK,heinzinger88,GXG,KKH,beu} or
{\it ab initio} pair potentials \cite{hannongbua00}
which all lead to essentially linear hydrogen bonds in the limit of
treating the ammonia dimer. 
A recent pioneering study \cite{DMT} investigated the 
structure of liquid ammonia through Car--Parrinello 
{\it ab initio} molecular dynamics \cite{CP,MH},
i.e.  
without resorting to any pre--determined (pairwise) potentials
but using the full many--body interactions as obtained by 
``on the fly'' DFT (BLYP) instead. 
%
%
It was found that the results of the N-N, N-H and H-H 
radial distribution functions are in better agreement with experimental 
results than previous ones relying on model potentials. 
Some important differences, however, still remained. 
For example, the results of this study were found to overestimate the hydrogen bonded 
structure as revealed by the N-H correlations and predict a narrower first peak of the 
N-N distribution when compared with the corresponding experimental results.
Also, a later such study \cite{LT} of liquid ammonia containing an ammonium or an amide solute
reported somewhat different results for the N-N and N-H correlations between the
bulk ammonia molecules compared to those of Ref.~\cite{DMT}.  
Besides, the nature of the 
hydrogen bonds in the liquid phase is not fully understood yet,
in particular in view of the experimental situation. 
As discussed
before, unlike water, the ammonia dimer has a bent hydrogen bond in its equilibrium
geometry. 
If this bent character of the dimer hydrogen bond has the same significant 
effects
on the liquid as in solid phases~\cite{nelmes96}
is still an unresolved issue, in particular
since all pair potentials by construction yield a quasi--linear hydrogen bond
as do existing density functionals.
%


The outline of this paper is as follows. 
In Section~II, we describe the details of the computational methodologies that we employ for the calculations
of the dimer and the liquid phase. 
In Section~III, we  present results for important dimer structures and their energies
based on highly accurate {\it ab initio} methods and compare these 
data to those obtained from a wide variety of currently available density functional families.
Section~IV focuses in detail on the development of the novel HCTH/407+ 
functional, on its performance in various systems, and on the improvement achieved on the dimer level. 
In Section~V, we compare the results of {\it ab initio} MD simulations of liquid 
ammonia using HCTH/407+ to those obtained from the widely used BLYP functional
for reference purposes, and to experimental data wherever available. 
We summarise and conclude in Section~VI.

\section{Methods and Technical Details}

\subsection{The HCTH Density Functional Family}

In this contribution, we develop a new generalised gradient approximation (GGA) functional
-- which will be termed HCTH/407+ --
that only involves the density $\rho$({\bf r}) and its gradient $\nabla\rho$({\bf r}). 
The latter has been shown to be 
a necessary ingredient for the description of hydrogen bonds within DFT,
see for instance Refs.~\cite{kari1,sim92,cpwatern,kari2,TBH,tHCTH,landman}.
To the contrary of
with the so--called hybrid functionals that involve ``Hartree--Fock exchange''
(such as e.g. B3LYP\cite{B3LYP}),
this advantage is not impaired by a prohibitive computational overhead 
relative to the local (spin) density approximation if used in the framework of 
plane wave pseudopotential calculations, 
see e.g. the arguments presented in Sect.~3.3 of Ref.~\cite{MH}.
Thus, we opted for
reparametrising the Hamprecht-Cohen-Tozer-Handy (HCTH) form, which has been described in
detail elsewhere \cite{HCTH93} and was originally used by Becke for a hybrid functional \cite{B97}.
In general, HCTH is a post-local spin density approximation (post- LSDA) functional,
meaning that it factorises the LSDA functional forms ($F_{LSDA}$)
\begin{eqnarray}
E_{xc} = \sum_{\gamma=x,c_{\sigma\sigma},c_{\alpha\beta}}\!\!\!\!\!\!\!E_{\gamma} = \sum_ {\gamma}
\sum_{q=0}^m c_{q,\gamma}
\int \!\!F_{LSDA,\gamma}(\rho_{\alpha},\rho_{\beta})
u_{\gamma}^q(x_{\alpha}^2,x_{\beta}^2) d{\bf r}
\enspace , 
\end{eqnarray}
here, $u_{\gamma}^q$ denotes the perturbation from the uniform electron gas if $c_{0,\gamma}$
is unity.
The sum over the integers $q$ going from zero to $m$ implies the power series defined through $u_{\gamma}$, with
$\theta_{\gamma\sigma}$ and $\eta_{\gamma\sigma}$ as fixed coefficients:
\begin{eqnarray}
u_{\gamma}^q(x_{\alpha}^2,x_{\beta}^2)=\left(\frac{\theta_{\gamma\sigma}
x_{\sigma}^2}{1+\eta_{\gamma\sigma}x_{\sigma}^2}\right)^q
\end{eqnarray}
The variable $x_{\sigma}$ is closely related to the reduced density gradient
\begin{eqnarray}
x_{\sigma}^2=\frac{(\nabla\rho_{\sigma})^2}{\rho_{\sigma}^{8/3}}
\enspace .
\end{eqnarray}
This functional form was previously employed
for reparametrisations \cite{HCTH120,HCTH407} resulting in a highly accurate
GGA functional, HCTH/407\cite{HCTH407}. Although the above form obeys
none of the exact conditions
for slowly varying densities \cite{Langreth},
and violates part of the scaling relations \cite{Levy},
it does obey the most crucial scaling relations and the Lieb-Oxford bound \cite{L-O,L-O2}
within the physically important region \cite{Shopping}.

When employing the form in Eqn. 1 up to fourth order in $m$, we obtain 15 linear coefficients
(because of having exchange, spin-like and spin-unlike correlation), which
are easily parametrised by minimising 
an error
function
$\Omega$, which is constructed out of the sum of three
components
\begin{eqnarray}
\Omega&=&\sum_{m}^{n_E} w_{m} (E_{m}^{exact}-E_{m}^{K-S})^2 + \sum_{l,X}^{n_G} w_{l,G} 
\left(\frac{\partial E_{l}^{K-S}}{\partial X}\right)^2\nonumber\\
&&+ \sum_{j,\sigma}^{n_v} w_{j,v} \int (v_{j,\sigma}^{ZMP} + k_{j,\sigma} - v_{j,\sigma}^{K-S})^2 \rho_{j,\sigma}^{2/3} d{\bf r} 
\enspace .
\end{eqnarray} 
The three sums
represent the mean-square deviations
from our reference data of
energies, gradients and exchange-correlation potentials for each molecule, respectively,
of the result of a Kohn--Sham density functional calculation 
(denoted by the superscript ``K-S'').
In the case of the total energies
reference data are
high--level quantum chemistry results that are denoted by ``exact''. 
In the second sum the exact gradients (at equilibrium geometry) are zero
by definition and thus do not appear in the formula. 
In the final term, we fit to the 
exchange-correlation potentials as determined by the Zhao-Morrison Parr(ZMP)-method~\cite{ZMP},
which are shifted 
by a constant $k$ because of the effects of the quantum-mechanical integer discontinuity.
The ZMP method has been proven to be an important
aspect of the fit~\cite{potential}.
All these contributions need to be 
balanced using weights $w$, which have
been determined and reported  previously~\cite{HCTH407}. 
The weights $w$ consist of a product of multiple partial weights making contributions 
for each molecule in order to ensure a balanced functional.

We also have to consider the molecular set for which the new functional was determined. 
In general, the
new HCTH/407+ functional was determined by fitting it to the ``407 set'' of molecules as
used for HCTH/407 \cite{www}. This set is similar to the G3 set of molecules \cite{G3}, with
added inorganic molecules, but a smaller proportion of large organic species which we considered
to be over--represented.
This original training set was  supplemented with data from the
ammonia dimer, including non-equilibrium structures,
as described in detail in Sect.~\ref{sec:hcth407+}.

\subsection{Quantum Chemical Calculations}

For the accurate determination of some points of the potential energy surface (PES) of the ammonia dimer,
we used the W2 method \cite{W2}. This is basically an extrapolation towards the full CCSD(T) basis set
limit including relativistic (but not Born-Oppenheimer) corrections. 
W2 energies are calculated at
the CCSD(T)/A$'$PVQZ geometry, this notation meaning an aug-cc-pVQZ basis set for nitrogen and a cc-pVQZ
basis set for
hydrogen \cite{PVXZfirst}. The $MOLPRO$ package \cite{MOLPRO} was used for these calculations. This method
is one of the most accurate standard {\it ab initio} methods currently available, with an average error
of less
than 0.5~kcal/mol for the G2-1 set\cite{G2-1} of molecules \cite{W2eval}. In particular, it is more
reliable and accurate than the G1, G2 and G3 methods (although computationally much more expensive)
\cite{W2eval,Mark}.

Two types of higher-order contributions were considered. Firstly, higher-order $T_3$ effects were assessed
by means of full CCSDT\cite{CCSDT} calculations using the $Aces II$ program system \cite{Aces}. Secondly, the
importance of connected quadruple excitations was probed by means of CCSD(TQ) \cite{CCSDtq} and BD(TQ)
\cite{BD,BDt} calculations using $Aces II$ and $Gaussian 98$ \cite{g98}, respectively.

For the DFT calculations, we employed the $Cadpac$ suite of programs \cite{Cadpac}, 
and assessed
a variety of GGA and meta--GGA functionals (namely 
BLYP\cite{BLYP},
PBE\cite{PBE},
HCTH/120\cite{HCTH120}, 
HCTH/407\cite{HCTH407}, 
$\tau$-HCTH\cite{tHCTH}, 
and
BP86\cite{P86};
we also tested but do not report results from
PW91\cite{PW91} and HCTH/93\cite{HCTH93})
and hybrids (B3LYP\cite{B3LYP}, 
B97-1\cite{HCTH93},
and $\tau$-HCTH hybrid\cite{tHCTH};
not reported are the results obtained from B97-2\cite{B97-2} and PBE0\cite{PBE0}). 
In case of the density functional calculations we used an A$'$PVTZ
basis set, which is reasonably close to the DFT basis set limit.

For the DFT values, which were obtained from using finite
basis sets without extrapolation as done in the W1/W2 methods,
we employed the
counterpoise (CP) correction~\cite{BSSE} to account for the basis set superposition error (BSSE).
In the fitting procedure, the TZ2P basis set\cite{TZ2P} was utilised in addition to the A$'$PVTZ basis set.

\subsection{{\em Ab Initio} Molecular Dynamics Simulations}

The {\it ab initio} molecular dynamics simulations 
were performed by
means of the Car-Parrinello method \cite{CP,MH}
and the {\tt CPMD} code~\cite{MH,cpmd}. A simple cubic box of 32
ammonia molecules with a box length of 11.229~\AA~ was periodically replicated
in three dimensions and the  electronic structure of the extended system
was represented by the Kohn-Sham formulation \cite{KS} of DFT within a plane wave basis.
The core electrons were treated via the atomic pseudopotentials of Goedecker
{\em et al.} \cite{GTH} and the plane wave expansion of the KS orbitals was truncated
at a kinetic energy of 70 Ry. 
A fictitious mass of $\mu$=875 a.u. was assigned
to the electronic degrees of freedom and the coupled equations of motion describing
the system dynamics was integrated by using a time step of 5 a.u. 
As usually done, see e.g. Refs.~\cite{cpwatern,tuck,Sprik},
the hydrogen atoms were given the mass of deuterium in order to 
allow for a larger molecular dynamics timestep
which also reduces the influence of (the neglected) quantum effects 
on the dynamical properties.

The {\it ab initio} MD simulations have been performed using the 
HCTH/407+ and BLYP~\cite{BLYP} functionals.  We
used the BLYP functional in addition to the new functional because the former  
has been shown to provide a good description of hydrogen 
bonded liquids such as water~\cite{cpwatern,Sprik,SP}, 
methanol~\cite{tsukada} and
also ammonia~\cite{DMT,LT}. 
Thus, it 
is worthwhile to compare the results of the
two functionals for various structural and dynamical properties of liquid 
ammonia. The initial configuration of ammonia molecules was generated by 
carrying out a classical molecular dynamics simulation using the empirical 
multi-site interaction potential of Ref.~\cite{GXG}.  Then, for  simulation
with the HCTH/407+ (BLYP) functional, we equilibrated the system for 8.75 
(10.1)\,ps at 273\,K in NVT ensemble using Nose-Hoover chain method and, 
thereafter, we continued the run in NVE ensemble for another 9.30 (9.34)\,ps 
for calculation of the various structural and dynamical quantities;
the average temperatures of these microcanonical runs were about 275 (252)~K.
We note that the size of the simulation box used in the present simulations 
corresponds  to the experimental density of liquid ammonia at 273~K which is 
$2.26 \times 10^{-2}$ molecules/~{\AA}$^3$~\cite{RNR}.

\section{The Ammonia Dimer: Structure and Energetics}

In order to briefly
validate
the W2 method for some representative hydrogen bonded systems, 
(HF)$_2$ and (H$_2$O)$_2$,
we compared the W2 results
to MP2 basis set limit corrections
applied to coupled cluster calculations\cite{Klopper2,Klopper1},
whereas the reference data was additionally empirically refined by
scaling the calculations to certain quantum energy levels.
Moreover, to get an estimate of the accuracy of the basis set
extrapolation, we compared with the more cost-effective alternative of W2,
which is W1 \cite{W2}. When calculated at the same geometry as W2,
it usually yields a fair estimate as to how well the basis set limit is obtained in the extrapolation.
In Table \ref{tab1}, the dissociation energies for the HF and H$_2$O dimers are compared to
best values from the literature, 
and displayed along with the results for the (NH$_3$)(H$_2$O) complex.
All the W1 and W2 results are within the stated
error margins of the reference data.
While the dissociation
energy of the hydrogen fluoride dimer is exactly the same, that of
water dimer is slightly 
lower than predicted
in reference \cite{Klopper2}. In the latter case, relativistic corrections
were
not included in their best estimate. This will reduce the energy somewhat; however we would expect
a geometry relaxation to lower the energy.
The authors of Ref.\cite{Klopper2} used a CCSD(T)/aug-cc-pVTZ
geometry and estimated the geometry relaxation, whereas the W2
calculation is carried out at an CCSD(T)/A'PVQZ
geometry. Nevertheless, the results show only 0.03 kJ/mol difference between the W1 and W2 method for
the HF dimer and 0.09 kJ/mol for the water dimer, indicating the high 
accuracy of these methods for such interactions. 
Finally, the (H$_2$O)(NH$_3$) dissociation energy is
predicted to be 26.8 kJ/mol by W2, compared to 27.4 kJ/mol 
determined by experiment \cite{Wurzberger}; 
unfortunately no high-level {\it ab initio} data were found in the literature
for this mixed dimer.

Let us now focus on the ammonia dimer. 
Five different structures on the dimer PES can be considered
important. The global minimum found on the CCSD(T)/A$'$PVQZ surface, as shown in Figure
\ref{fig1}c, has an HNN angle of 19.9$^{\circ}$ with eclipsed hydrogens and C$_s$ symmetry. Its staggered
counterpart, 
displayed in Figure \ref{fig1}d, has an HNN angle of 13.1$^{\circ}$. The two completely linear
structures with the HNN angle being 0$^{\circ}$ (Figures \ref{fig1}a and \ref{fig1}b)
are salient points and the C$_{2h}$ structure (Figure \ref{fig1}e) 
is a transition state.
In Table \ref{tab2}, 
W1 and W2 results for the five structures are summarised, with column two and four (with BSSE) 
corresponding to the regular W2 method without the cp-correction. Because of
the low barrier between the C$_s$ minimum and
the C$_{2h}$ structure, we additionally calculated a counterpoise-corrected W2 barrier. As to
the accuracy of the electron correlation methods underlying the W2 method, we can make the following
observations:
\begin{itemize}
\item
quasiperturbative connected triple excitations, i.e., the CCSD(T) -- CCSD difference, 
only contribute 0.5 cm$^{-1}$ to the C$_{2h}$ barrier with CP correction and 0.7 cm$^{-1}$ without;

\item this has also been verified by CCSD(TQ) [BD(TQ)] calculations, where the contribution of the
perturbative
connected quadruples lowers the barrier by 0.2 cm$^{-1}$ [0.3 cm$^{-1}$] with the A$'$PVDZ basis set;

\item higher-order connected triples 
(i.e., CCSDT-CCSD(T))
additionally lower the barrier by 0.2 cm $^{-1}$ when using
A$'$PVDZ. This leads to the conclusion that not only has basis set convergence been achieved, but also
that higher-order excitations 
beyond CCSD(T)
will contribute less than 1 cm$^{-1}$;

\item further optimisation of the C$_s$ geometry at the MP2/A'PV5Z level lowers the binding energy by
only 0..3 cm$^{-1}$ (0.004 kJ/mol) relative to MP2/A$'$PV5Z//MP2/A$'$PVQZ;

\item we do not expect relaxation of the geometry at the core-valence level to make a significant
contribution, since even for the monomer the core-valence correction
to the geometry is very small \cite{core}.
\end{itemize}
Still, the difference between the BSSE-corrected and uncorrected results is rather large. Based on
these results, we suggest that the C$_s$ to C$_{2h}$ energy difference is 3.5 $\pm$ 3 cm$^{-1}$,
and we would expect the barriers between the linear and minimum geometries to be equally well
described by W2.
For the slightly improved CCSD(T) geometry, by correcting CCSD(T)/A$'$PVQZ - MP2/A$'$PVQZ + MP2/A$'$PV5Z,
we predict the HNN angle to be 20.7 degrees.
As for the C$_{2h}$ barrier, the interaction energy is raised by 0.002 kJ/mol when going from
CCSD(T) to CCSDT, showing that the
quasiperturbative triples treatment (and the quadruples estimation 
as well) is quite accurate. Using an A$'$PVDZ basis set, the quasi-perturbative quadruples treatment lowers
the binding energy by 0.092 kJ/mol for the
coupled cluster and 0.087
kJ/mol when using the Brueckner-Doubles method.
With the geometry relaxation contributing less than 0.005 kJ/mol at MP2 level and the estimate of the
higher order excitations, we predict the dissociation energy to be 13.1 $\pm$ 0.2 kJ/mol, since the
main error is likely to come from the quadruple excitations.

The results obtained using a selection of currently published functionals are shown
in Table \ref{tab3}. For comparison, the CCSD(T), CCSD, MP2 and HF values are given. It should be
noted, however, that
BSSE for the coupled-cluster values is more than twice as high as for DFT (0.9 compared to 0.4
kJ/mol), and therefore we would expect the DFT calculations to be closer to their respective basis set
limits than the {\it ab initio} methods. Nevertheless, their BSSE is still larger than
that at the HF level.
As expected,
the CCSD(T) and W2 numbers are reasonably close to each other. Most density functional methods reproduce
the dissociation energy
as listed in the first column, reasonably well. In
general, the GGA and meta-GGA functionals tested give a good description of this interaction
underestimate and PW91 overestimating the dissociation energy.
The tested
hybrid functionals are even more accurate, yet all of them underestimating the dissociation energy.

However, if we consider the 
energy difference between the staggered and eclipsed conformation of the NH$_3$-dimer
(see the data in the second column of Table \ref{tab3}), 
it becomes obvious that 
available functionals are
incapable of reproducing the effect of a bent hydrogen bond.
It is found that density functionals typically underestimate 
the difference between the local and global minima by at least a factor of two.
We note here that
functionals other than those listed in our tables 
(such as e.g. PW91, B97-2, PBE0)
yield similar results
which are therefore not displayed.
%
On the other hand, all other wavefunction-based {\it ab initio} methods considered are able to
reproduce this effect. This becomes even more apparent when considering the C$_{2h}$ transition state.
Here, the coupled-cluster differences are close to 3 cm$^{-1}$, whereas HF renders 24 cm$^{-1}$ and the DFT
values are between 60 and 95 cm$^{-1}$ with the exception of HCTH/407 (27 cm$^{-1}$) and the
$\tau$-HCTH functionals (100 and 140 cm$^{-1}$). In turn, the other
barrier is underestimated by 30-60\% and both linear structures are stabilised relative to the
global minimum.
The same trends are seen in Table \ref{tab4}, where the HNN angles of the two minimum structures
are shown. Here, available density functionals 
underestimate the bending of the hydrogen bond by almost a factor of two for the
global minimum and by about 30\% for the local (staggered) minimum, although the hydrogen bond
distance itself is
reasonably well
reproduced. Both $\tau$-HCTH functionals, while they perform well with their
large 'training' fit set, are the worst performers here. On the other hand, HCTH/407 is the
only functional which is somewhat capable of reproducing the effect, outperforming Hartree-Fock.

In summary, although standard density functionals
do yield a correct dissociation energy and a
slightly nonlinear hydrogen bond, they completely overestimate 
the barrier towards the symmetrical C$_{2h}$ structure,
thereby preferring a much more linear hydrogen-bonding 
configuration than our accurate reference methods.
Although the energy differences might be viewed as intrinsically too small to be accurately
rendered by DFT, the fact
that even HF gives better results is somewhat disturbing. Nevertheless, the observation that
Hartree-Fock and HCTH/407 are somewhat
capable of properly predicting the structure and energetics of the ammonia dimer system suggests that
the problem at hand is not beyond the grasp of density functional methods
as such.

\section{The HCTH/407+ Functional: Construction and Performance}
\label{sec:hcth407+}

Considering that standard density functionals 
were shown to be unable to describe the energy difference
between linear
and bent hydrogen bonds 
and do not capture the large bending angle of the hydrogen bond,
the most obvious remedy would be to change the functional form itself. However,
if all tested functionals give such a bad description, this is unlikely to solve the problem. The
culprit in this case, therefore, has to be the set of molecules which the 
HCTH functionals are fit to,
insofar as only linearly hydrogen-bonded systems have been included. Nevertheless, including the minimum
structure and energetics of the ammonia dimer
determined in the previous section into the set probably only partly solves this problem.
The main problem, as established before, is the energy difference between the linear and bent hydrogen
bonds. In addition, the TZ2P basis set commonly used in the fit set might not be appropriate. Even when
employing the A$'$PVTZ basis set mentioned above, the basis set superposition error
for the dimerisation energy is 29 cm$^{-1}$,
contributing 8 cm$^{-1}$ to the energy difference between the minimum and symmetric structure when using
the B97-1 functional. In comparison, using the TZ2P basis set, the BSSE is 92 cm$^{-1}$, contributing 
33 cm$^{-1}$ to the energy difference. Clearly, the TZ2P basis set is inappropriate for this problem.
The above numbers indicate that a counterpoise procedure has to be employed within the
fit in order to get a good estimate. Note
that even when using an extrapolation to the
full basis set limit (in the case of W2), the difference between the counterpoise corrected and
uncorrected values is already 3 cm$^{-1}$.

All of the above issues were considered in the fit.. Thus, we fit to all points shown in Figure
\ref{fig1} employing an A$'$PVTZ basis set with counterpoise correction in the fit. This is the
first time we are fitting to non-equilibrium structures, and we only fit to their energy differences
and exchange-correlation potentials of these. Due to the energy difference of only
a couple of wavenumbers, an additional weight has to be employed for any change in the fit to be
noticeable. Thus, equation (3) now reads
\begin{eqnarray}
\Omega&=&\sum_{m}^{n_E} w_{m} (E_{m}^{exact}-E_{m}^{K-S}-E_{m}^{BSSE})^2 + \sum_{l,X}^{n_G} w_{l,G} \left(\frac{\partial E_{l}^{K-S}}{\partial X}\right)^2\nonumber\\
&&+ \sum_{j,\sigma}^{n_v} w_{j,v} \int (v_{j,\sigma}^{ZMP} + k_{j,\sigma} - v_{j,\sigma}^{K-S})^2 \rho_{j,\sigma}^{2/3} d{\bf r}
\end{eqnarray} 
with
\begin{eqnarray}
E_{m}^{BSSE} = E_{m}^{ghost,donor} + E_{m}^{ghost,acceptor} - E_{m}^{monomer,donor} - E_{m}^{monomer,acceptor}
\enspace .
\end{eqnarray}
Since we do not want to fit to a given value of basis set superposition error, it is included
in $E_{m}^{K-S}$. The weights $w_{m,l,j}$ now also include the additional weights for the
potential energy surface with the other weights defined as before \cite{HCTH407}. Thus, 
$w_{m}$ is composed of
\begin{eqnarray}
w_{m,PES}= 750\times{\rm weight}_{confidence}\times{\rm weight}_{atom}\times{\rm weight}_{dimer}\times{\rm weight}_{PES}
\enspace .
\end{eqnarray}
When determining the weights for the new members of the fit set, we need to consider both their
A$'$PVTZ optimised energies as well as their A$'$PVTZ single-point energies at the CCSD(T)/A$'$PVQZ optimised
geometries. The latter is necessary since we can only fit to single points. As a result of
the small energy differences, the 
frozen-geometry single-point energies and optimised energies differ by a significant
amount.
The development procedure is similar to the one employed for the HCTH/407 functional, and the set 
includes numerous atomisation energies, proton affinities, ionisation potentials, electron affinities,
and dissociation energies of hydrogen-bonded dimers and transition metal complexes \cite{HCTH407}. First,
we fit to a subset of the 407 molecules in order to get close to the global minimum, then we fit to the full
set. The latter subset (small set) is the G2-1 set plus two hydrogen-bonded systems (H$_2$O)$_2$ and (HF)$_2$,
 and the additional structures on
the potential energy surface, making 153 systems in all (the 147 of the HCTH/147 functional plus the
five PES
points, in addition to the NH$_3$ dissociation energy at the TZ2P basis). For each point on the PES, the
counterpoise correction was employed. The results using the newly determined functionals as
a function of the weights have been determined in both the first fit (to 153 systems) and the second fit
(to 413 systems). Thus, we obtained new functionals with weights
ranging from 0/0 (which corresponds
to the HCTH/407 functional)
to 200/200. The first weight is the one used in the subset, while the second
corresponds to the large fit.
The importance of the weights in the first
fit can be seen from the differences by using
weight pairs of 30/80 and 80/80
- they differ by about a
factor of two in the energy difference between the staggered and eclipsed geometry. Hence, the minimum obtained
by the second fit will
of necessity be close to the global minimum obtained by the first one. 
When increasing these weights,
we approach the reference values as expected, with the 200/200 functional rendering the closest values to
the reference method. While not explicitly included in the fit, the BSSE also increased from 0.444 kJ/mol
to 0.866 kJ/mol. This resembles the tendency of the functional to have a larger intermolecular distance
than the reference methods when increasing the weights, hence entering a regime where the BSSE increases
more rapidly than for the optimised structures. In comparison, the largest BSSE observed by the
functionals tested in Table \ref{tab3} is 0.48 kJ/mol. However, for all the calculations in the fit we
have performed only DFT single-point energy calculations at the CCSD(T) optimised geometries.

Another, more appropriate, point of comparison
consists of the optimised structures of all the newly obtained
functionals, as done in Table \ref{tab3} for the contemporary functionals.
Since the energy differences in general are very small, these results differ
significantly from the ones obtained by the single-point energies calculated at the
coupled-cluster geometries. In addition, the C$_{2h}$ structure becomes
the global minimum for the functionals with a weight of 80 or larger in the first fit. Based on these
results, the 30/80 functional appears to be the most reliable for this interaction, and hence will be
denoted as
the new HCTH/407+ functional. This new functional (compare to Table \ref{tab3}) now renders a 
dissociation energy of 13.18~kJ/mol 
%
for the ammonia dimer, in much closer agreement to the reference values than any other
method employed. The difference between the C$_s$ minimum and the C$_{2h}$ structure is 4.0 cm$^{-1}$ for HCTH/407+,
compared to our best estimate of 3.5 cm$^{-1}$ in Table \ref{tab2}. The staggered C$_s$ structure lies 18 cm$^{-1}$ above
its eclipsed counterpart, which is about five wavenumbers higher than our 
best
estimate but still within its error bars.
The difference to the linear eclipsed and linear staggered structures are slightly underestimated
at 51.5 and 53.5 cm$^{-1}$ respectively, but this is still a vast
improvement over the other functionals. In Figure \ref{fig2}, we 
compare the energies of the new HCTH/407+ functional to W2 and 
one of the most commonly used GGA functionals (BLYP).
The BSSE for the relaxed C$_s$ structures
does not increase as rapidly as for their CCSD(T) optimised
counterparts. It increases from 0.43 (HCTH/407) to 0.45 (HCTH/407+) to 0.61 kJ/mol (weight of 200),
and does not yield as unreasonably large values as the single-point calculations at the CCSD(T)
geometries. The hydrogen bond lengths of the C$_s$ and C$_{2h}$ structures increase 
with larger weights, emphasising the importance of this analysis. Generally, 
all HCTH functionals are found to slightly overestimate this hydrogen
bond length. The HNN angle of the HCTH/407+
functional is 17 $^{\circ}$ for the minimum geometry. This is within the accuracy that can be expected
from contemporary density functionals ($\pm$ 2$^{\circ}$), showing a significant 
improvement over all other
density functionals in Table \ref{tab4}. 
The results for all functionals obtained with different
weights (from which we determined the HCTH/407+ functional) are given in the supplementary material.
\cite{supplementary}

The parameters 
which define 
the HCTH/407+ functional, compared to HCTH/407, are given in Table \ref{tab5}. As
might be expected, they differ only slightly from those for the standard HCTH/407 functional;
only the correlation and higher-order coefficients have significantly been affected by the
change.. Nevertheless, the new functional should now be able to describe non-directed hydrogen bonds
better 
than HCTH/407.
%
On the level of the ammonia dimer, the improvement with respect to other
density functionals becomes clearly visible in Fig.~\ref{fig2};
see Table~\ref{tab3} for the corresponding numbers.
It is seen that  HCTH/407+ describes the relative stability of all isomers
much better with respect to the W2 reference data than BLYP, 
which was selected as a prominent representative of the GGA family.
Most importantly, the dramatic failure of BLYP to capture the
stability of the cyclic C$_{2h}$ isomer 
relative to all other isomers is cured.
According to HCTH/407+, the two $C_s$ structures as well as the
$C_{2h}$ structure are essentially degenerate, which is in accord with the reference data.. 
Apart from the energetics, also the structure of the 
ground state of the ammonia dimer, i.e. the eclipsed $C_s$ structure, is dramatically improved,
see Table~\ref{tab4} for details. 
In particular the hydrogen bond angle HNN is now much larger and thus closer 
to the reference value than {\em any} other density functional method; 
note that also the functionals PW91, B97-2, and PBE0 were considered.
The same is true for the HNN angle in the staggered variant of the $C_s$ symmetric structure.
However, it is also clear from this table 
that the N$\cdots$H  distance
is clearly overestimated: HCTH/407+ yields a distance of about 2.5~\AA~
instead of around 2.3~\AA~ as required.
This seems to be typical for the HCTH family as both HCTH/120 and
HCTH/407 yield similarly large values exceeding 2.4~\AA~
(and HCTH/93 leads to 2.614~\AA).
In addition, the same trend of producing too long hydrogen bonds 
is found for the $C_{2h}$ structure. 
This deficiency is corrected by both $\tau$-HCTH and $\tau$-HCTH hybrid
functionals.  
Here it should be noted that both BLYP and PBE describe the
hydrogen bond length quite well, however at the expense of
making it much too linear and grossly disfavoring the cyclic structure energetically.

Furthermore, 
this significant improvement in describing 
non-directed hydrogen bonds, however, comes at the expense of
a slightly increased error for the other molecules in the fit set. 
Table \ref{tab6} shows the errors of
the HCTH/407 and HCTH/407+ functionals to the fit set, summarising these as RMS energy and geometry errors,
together with the errors for the energies, geometry shifts and frequency shifts of nine
hydrogen-bonded systems. The values for B3LYP, BLYP and BP86 are included for comparison. The
hydrogen-bonded systems are the (HF)$_2$, (HCl)$_2$, $\rm (H_2O)_2$, (CO)(HF), (OC)(HF),
(FH)(NH$_3$), (ClH)(NH$_3$),
$\rm (H_2O)(NH_3)$ and $\rm (H_3O^+)(H_2O)$ complexes, but not the NH$_3$ dimer itself. The
performance of other functionals over all the sets has been discussed elsewhere \cite{TBH,HCTH407,tHCTH}.
Overall, errors for the HCTH/407+ functional are quite similar to HCTH/407, which in turn means
that they are still excellent compared to first-generation hybrid functionals like B3LYP and
especially GGA functionals like BLYP or BP86. Although we expect the geometries to be still
slightly worse than B3LYP (although the gradient error is comparable), the energetic properties
are better described, as are the hydrogen bonds. Whereas the errors of the HCTH/407+ functional for
the fit set have barely changed compared to HCTH/407, the accuracy of the dissociation energies of
the nine directed hydrogen bonds has actually decreased by about 20\%. Nonetheless, the accuracy of the
frequency shifts of the H-X bond stretches has increased,
indicating that we 
seem to have 
improved
the description of the potential energy surface of these complexes. 
%

\section{Liquid Ammonia: Structure, Dynamics, and Hydrogen Bonding}

\subsection{Radial Distribution Functions}


We have calculated the nitrogen-nitrogen, hydrogen-hydrogen and nitrogen-hydrogen
radial distribution functions (RDF) from the atomic trajectories generated by 
{\it ab initio} molecular dynamics 
simulations;
note again that we used the mass of the deuteron instead of that of the proton 
but we still use the symbol H for convenience.
The results for both HCTH/407+ and BLYP functionals are
shown in Fig.~\ref{fig3}. In this Figure, we have also included the experimental results
\cite{RNR,Soper}
of the RDFs of liquid ammonia at 273~K;
note that the underlying experimental technique is based on the assumption
that the structure of all isotopically substituted systems is identical. 
For the 
nitrogen-nitrogen RDF, it is seen that at the short distances the BLYP $g_{NN}(r)$ 
is in better agreement with experimental results. The HCTH/407+ peak is located at 
a distance of about 0.15~\AA~ larger than the experimental peak position whereas
the BLYP peak appears at a somewhat shorter distance.  
We note in this context that 
the HCTH/407+ functional gives an overestimated equilibrium N-N distance for the 
isolated dimer compared to the reference CCSD(T) result as can be inferred from Table~\ref{tab4}.
The  larger N-N distance  
seems to be translated to the liquid phase configurations and we observe 
a somewhat larger most probable N-N distance for the HCTH/407+ functional than what is 
found in the experiments.  Still, the
overall shape of the first peak is represented reasonably well by both the functionals.
When the RDFs are integrated up to their first minimum to obtain the 
coordination number in the first solvation shell, we obtained the values of 13.2
and 12.1 for HCTH/407+ and BLYP functionals
compared to an experimental value
of 12.75. Clearly, the agreement with experimental results
is found to be rather good.

Considering the results of the H-H and N-H RDFs, the positions of the
calculated intramolecular H-H and N-H peaks for both the functionals agree
rather well with
experimental results as expected. For example, the liquid phase N-H bond 
length and intramolecular H-H distance for 
the HCTH/407+ (BLYP) functionals are 1.03 (1.03)~\AA~
and 1.63 (1.65)~\AA~ which can be compared with the experimental values of 1.04~\AA~
and 1.60~\AA, respectively. Also, these intramolecular distances in the liquid phase
are very close to their gas phase values. This was also noted earlier 
for the BLYP functional \cite{DMT}. Because of the quantum disperson effects which are
present in real liquid but not in the current simulated systems,  the experimental
intramolecular peaks are found to be broader than the theoretical peaks. The
somewhat higher value of the experimental H-H correlation at the first minimum at 
around 2~\AA~ can also be attributed to such quantum dispersion effects. 

The intermolecular
H-H peaks  are  found to be well represented by both
functionals. 
The intermolecular 
part of the N-H RDF  shows the presence of both H-bonded and non H-bonded
H atoms in the first solvation shell. The shoulder up to about 2.7~\AA~ can be
attributed to hydrogen bonded H atoms,
while the more pronounced peak at the larger
distance at around 3.75~\AA~ corresponds to the hydrogen atoms in the solvation
shell which are not hydrogen bonded. Again, both functionals are found to describe
well the non-hydrogen bonded first solvation shell peak when compared with the
corresponding experimental result.  For the hydrogen bonded part, however, the HCTH/407+ 
result is found to
exhibit a less pronounced shoulder than what is found in 
the experiments and for the BLYP functional.
This is probably a result of the slight overestimation of the hydrogen bond
length by HCTH/407+ already on the level of the dimer, see
Table~\ref{tab4}.
Considering both hydrogen bonded and
non-hydrogen bonded H atoms, a sphere of radius 5.3 (5..2)~\AA~ around a central
N atom is found to contain 34.1 (34) intermolecular H atoms for
the HCTH/407+ (BLYP) functional.
The radius of the sphere is set to 5.3 (5.2)~\AA~ because the first minimum of the 
intermolecular N-H RDF is located at this distance. The corresponding radius
for the experimental N-H RDF is 5.0~\AA, yielding the experimental value
of 34.8 for the above coordination number. A more detailed analysis of the distribution 
of hydrogen bonds in liquid ammonia is given in the following subsection, and the 
dynamical aspects of hydrogen bonds are considered in Sect.~\ref{sec:hbdyn}.

\subsection{Distribution of Hydrogen Bonds}
\label{sec:hbstruc}

The analysis of the hydrogen bond (HB) distribution among ammonia molecules is
based on a calculation of the fractions $P_{n}$ of ammonia molecules that
engage in $n$ hydrogen bonds and the average number of hydrogen bonds that 
an ammonia molecule donates and accepts~\cite{noteHB}. 
%
Following previous work on liquid ammonia,
water and other hydrogen bonding liquids \cite{DMT,GXG,KKH,GSR,LC,XB,SNS,AC,BPB,FHM},
we have adopted a geometric definition of the hydrogen bonds where two ammonia molecules 
are assumed to be hydrogen bonded if they satisfy the following configurational criteria 
with respect to N-N and N-H distances
\begin{eqnarray}
R^{(NN)}&<&R^{(NN)}_{c} \ , \nonumber\\
R^{(NH)}&<&R^{(NH)}_{c}
\end{eqnarray}
and the values of these distance cut-offs are usually determined from the
positions of the first minimum of the corresponding (intermolecular)
RDFs. We have used a similar procedure where the N-N RDF gives a value of R$^{NN}_{c}$
=5.25 (5.1)~\AA~ for the N-N cut-off for HCTH/407+ (BLYP) functionals. However, 
the procedure can not be applied unambiguously to determine the N-H cut-off 
because no clear minimum
that separates the hydrogen bonded and nonbonded H atoms is found in the N-H RDF.
 In the latter case, the shoulder which is assigned
to hydrogen bonded H atoms merges into the broad and more intense peak corresponding
to the H atoms of the first solvation shell which are not hydrogen bonded. Since
this shoulder
-- which corresponds to the hydrogen bonded H atoms -- extends up to about
2.7~\AA~ in the experimental and BLYP
RDFs we have  taken this distance of 2.7~\AA~ as the
cutoff N-H distance. For HCTH/407+ also, we have used the same cut-off of the N-H distance 
although  the hydrogen bonded shoulder is less pronounced for this functional.
Still, variation of $R^{NH}_{c}$ within reasonable bounds does not change
qualitatively the result of this analysis.

In Fig.~\ref{fig4}, we have shown the distribution of donor and acceptor hydrogen bonds 
for both HCTH/407+ and BLYP functionals. In this figure, panels (a) and (b) show
the fraction of ammonia molecules that accept or donate $n$ hydrogen bonds
and panel (c) shows the fraction of hydrogen atoms that donate $n$ hydrogen bonds. 
The majority of the ammonia molecules is found to accept one hydrogen bond and to 
donate one. Also, the fraction of hydrogen atoms donating two or more hydrogen bonds is 
found to be negligibly small,
which means that the bifurcated or multifurcated hydrogen bonds 
are practically absent in liquid ammonia.
The average number of donor hydrogen bonds per ammonia molecule, which
is also equal to the average number of acceptor hydrogen bonds per molecule, is found 
to be 0.93 (1.34) for the HCTH/407+ (BLYP) functional. Given that there are as
many 
as about  13 ammonia molecules in the first solvation shell, we conclude that 
non--hydrogen--bonded interactions, i.e. packing or steric effects, 
are crucial in determining the solvation behavior of ammonia in the liquid state. 
This is in 
sharp contrast to the case of liquid water where the solvation structure of a  water 
molecule is determined primarily by hydrogen bonding interactions \cite{FHS}.

The ammonia dimer has a strongly bent hydrogen bond in its equilibrium geometry and
it would be interesting to
investigate whether the hydrogen bonds remain bent in the 
liquid phase. We have
carried out such an analysis by calculating the distribution 
function of the cosine of the hydrogen bond angle $\theta$, 
which is defined as the  N$\cdots$ N--H
angle of a hydrogen bonded pair and the results are shown in Fig..~\ref{fig5}a for both
the functionals. It is seen that, for both the functionals, the probability 
distribution  $P(\cos{\theta})$ is peaked at $\cos{\theta}=1$ 
indicating that the {\em preferred hydrogen bond geometry is linear in the liquid phase}, 
see Ref.~\cite{cone-corr}.
The peak is somewhat more sharp and narrower for the BLYP functional, which 
means a stronger 
preference for the linear hydrogen bonds for this functional as compared to that for
the HCTH/407+ functional. This preferential presence of linear hydrogen bonds in the
liquid phase is indeed an interesting result given that the dimer has a bent
hydrogen bond in the gas phase.
We have also investigated the distribution of N$\cdots$ N--H angle for 
H atoms that belong to nearest neighbours but
are not hydrogen bonded, i.e. the N$\cdots$H distance is greater than 2.7~\AA. The results of
these angular distributions are shown in Fig.~\ref{fig5}b for 
2.7~\AA~ $<$ R$^{(NH)}$ $<$ 3.7~\AA~
and in Fig.~\ref{fig5}c for 3.7~\AA~ $<$ R$^{(NH)}$ $<$ 4.7~\AA. A wide range of values
are found for the N$\cdots$ N--H angle with very little or no preference for a particular 
orientation. 
These findings are indeed consistent with the above analysis 
that the first solvation shell in liquid ammonia is dominated by simple packing 
rather than by directional hydrogen bonding.

\subsection{Self Diffusion and Orientational Relaxation}

The translational self diffusion coefficient of an ammonia molecule 
in the liquid state is 
calculated from the long-time limit of the mean-square displacement (MSD) 
\begin{equation}
D=\lim_{t\rightarrow \infty} \frac{<[r(t)-r(0)]^{2}>}{6t}\ ,
\end{equation}
where $r(t)$ is the position of the center of mass of a molecule at time $t$ 
and  the average is carried out over the 
time origin for autocorrelation
and over all the molecules
as usual. 
The results of the diffusion coefficients, which are obtained
from a least square linear fit of the simulation data excluding the
initial ballistic regime up to 
0.5\,ps,  are included in 
Table (VII). This Table also contains the experimental values
of the diffusion coefficients of 
deuterated liquid ammonia 
ND$_{3}$ at 275 and 252\,K
that are found from Eq.(2) of Ref.\cite{OPS},
which was shown to
fit
the experimental data over a rather wide temperature range
very well. The agreement 
of the calculated diffusion coefficients with the experimental result
is found to be reasonably good for both
functionals. 
In particular it is gratifying to see that the experimental value
is {\em underestimated} 
because quantum effects on nuclear motion, which are neglected
in the present study, tend to increase the diffusion in a liquid.

The self-orientational motion of ammonia molecules is analyzed by calculating the 
orientational time correlation function~\cite{md-ana}, $C_{l}^{\alpha}(t)$, defined by 
\begin{equation}
C_{l}^{\alpha}(t)=\frac{<P_{l}(e^{\alpha}(t)\cdot e^{\alpha}(0))>}
{<P_{l}(e^{\alpha}(0)\cdot e^{\alpha}(0))>} \ ,
\end{equation}
where $P_{l}$ is the Legendre polynomial of
order $l$ and $e^{\alpha}$ is the 
unit vector which points along the $\alpha$-axis in the molecular frame. Here,
we have studied the time dependence of $C_{l}^{\alpha}(t)$ for $l=1,2$ and for 
three different $e^{\alpha}$, the unit vectors along the molecular dipole
axis, an N-H bond and an intramolecular H-H axis. The geometric dipole vector
of an ammonia molecule is calculated by assigning partial charges to N and
H atoms as given by the classical model of Ref.\cite{GXG}. For the orientational 
relaxation of unit vectors along the N-H and H-H axes, the 
results are averaged over three such intramolecular axes of each type. After
an initial transient non-exponential decay, the relaxation becomes diffusional
and $C_{l}^{\alpha}(t)$ decays exponentially. The orientational correlation 
time, $\tau_{l}^{\alpha}$, is defined as the time integral of the orientational 
correlation function 
\begin{equation}
\tau_{l}^{\alpha}=\int_{0}^{\infty} dt \ C_{l}^{\alpha}(t)\ . 
\end{equation}
We note that the orientational relaxation of molecular vectors containing H 
atoms are usually measured by NMR relaxation experiments, which yield the 
Fourier transform of a correlation function for $l=2$
such that the integrated
correlation  time $\tau_{2}^{\alpha}$ is measured. 

In Fig.~\ref{fig6}, we depict the results of the orientational correlation 
function $C_{l}^{\alpha}(t)$ for $l=1,2$ and $\alpha$=dipole, N-H and H-H. The 
corresponding results of the orientational correlation times are included in 
Table (VII). We have also included the experimental result of $\tau_{2}$ of 
deuterated ammonia 
$ND_{3}$ as given by the fitted Arrhenius law of Ref. \cite{HZZ}. Since this 
experimental relaxation time was obtained by NMR relaxation experiments, it is 
likely to correspond to the relaxation of N-H/H-H vectors rather than the dipole 
vector. The BLYP functional is found to predict a relatively slower
orientational relaxation. Considering the quite small system size and
short simulation trajectories, both functionals are found to
describe the single-particle rotational dynamics of liquid ammonia reasonably well.

\subsection{Dynamics of Hydrogen Bonds}
\label{sec:hbdyn}

The key dynamical quantity in the context of hydrogen bond dynamics is the
mean hydrogen bond life time, $\tau_{HB}$. 
The fast librational motion of ammonia molecules makes
apparent breaking and 
reformation of a hydrogen bond a very fast process and,  
depending on how these fast transient events are taken into account, one can obtain 
different values for the  hydrogen bond (HB) lifetime. 
%
The rotational motion of ammonia molecules is also a rather fast process and can 
also contribute to the short--time dynamics of hydrogen bonds.  
For a
particular chosen definition of a breaking event, one can adopt a 
direct counting method \cite{FHM,HFM,RDM} or a time correlation function 
method \cite{LC,XB,SNS,AC,BPB,RAP} to calculate the lifetime. 
In the present work, we have adopted the time correlation
function approach which allows us to calculate the hydrogen bond lifetime in both
cases: (a) The breaking of hydrogen bond occurs due to fast librational motion even
though it may be a transient event and not a `true' breaking event and (b) these 
fast librational breaking events are ignored and only
rotational and translational diffusional breaking on a longer time scale is considered
as the `true' breaking of a hydrogen bond. 
Also, in the present context, it is implied that `fast librational
breaking' also includes the effects of fast rotational motion that might
contribute to the short time dynamics of the hydrogen bonds.

In order to obtain the hydrogen bond lifetime for the two scenarios mentioned above,
we calculate three time correlation functions: one continuous and two
intermittent hydrogen bond correlation functions \cite{LC,XB,RAP}.
Before we define these time
correlations, we first define two hydrogen bond population variables $h(t)$ and $H(t)$:
$h(t)$ is unity when a particular tagged pair of ammonia molecules is 
hydrogen bonded at time $t$, according to the adopted definition
from Sect.~\ref{sec:hbstruc} 
and zero otherwise.
The function $H(t)$ is unity if the tagged pair of ammonia molecules remains 
continuously hydrogen bonded from $t=0$ to time $t$ given that the bond
was formed for the last time at $t=0$, and it is zero otherwise.
We define the continuous hydrogen bond time correlation function $S_{HB}(t)$ as 
\begin{equation}
S_{HB}(t)=<h(0)H(t)>/<h> \ ,
\end{equation}
where $<\cdots>$ denotes an average over all hydrogen bonds  that
are created at $t=0$. Clearly, $S_{HB}(t)$ describes the probability that 
a hydrogen bonded ammonia pair, which was created at $t=0$, remains continuously
bonded upto time $t$. It differs from the continuous hydrogen bond correlation function
of Rapaport \cite{RAP} in that we
apply the condition of bond formation at
time $t=0$.  The mean hydrogen bond lifetime $\tau_{HB}$ can then be calculated
from the time integral
\begin{equation}
\tau_{HB}=\int_{0}^{\infty}S_{HB}(t)\,dt \ .
\end{equation}
In Fig.~\ref{fig7}a, we have shown the decay of $S_{HB}(t)$ for both functionals
and the corresponding results of $\tau_{HB}$ are included in Table VII.
The hydrogen bond lifetime is found to be about 0.1\,ps for both functionals. 
We note that both fast 
librational and slower diffusional motion can contribute to the decay of 
$S_{HB}(t)$. Since the librational motion occurs
on a faster time scale and 
since the correlation function $S_{HB}(t)$ does not allow any reformation 
event, the dynamics of $S_{HB}(t)$ primarily
reveal the dynamics of hydrogen bond 
breaking due to fast librational motion and hence the lifetime that is obtained 
from Eq.(12) corresponds to the average time over which a hydrogen bond survives
before it `breaks' due to librations. 
Again, this short time breaking may also include contributions from fast 
rotational motion of ammonia molecules which also occurs at subpicosecond 
time scale.

A different way to analyze the hydrogen bond dynamics is to calculate
the intermittent hydrogen bond correlation function \cite{LC,RAP}
\begin{equation}
C_{HB}(t)=<h(0)h(t)>/<h>
\enspace ,
\end{equation}
where the average is now over all hydrogen bonds that were present at $t=0$. 
Note that the correlation function $C_{HB}(t)$  does not depend on the
continuous presence of a hydrogen bond. It describes the probability that
a hydrogen bond is intact at time $t$, given it was intact at time zero,
independent of possible breaking in the interim time. Clearly, bonds which
were briefly `broken' by fast librational motions would continue to contribute
to the correlation function at later times and this leads to a much slower 
decay of $C_{HB}(t)$ at longer times. The results of $C_{HB}(t)$ are shown in 
Fig.~\ref{fig7}b which clearly shows a
two-phase relaxation of this correlation function.  
The initial fast relaxation corresponds to
rapid breaking of hydrogen bonds due to librational
motion and the slower relaxation after this initial transient decay corresponds
to the breaking and reformation of hydrogen bonds due to both rotational and
translational diffusion of ammonia molecules. 

After a hydrogen bond is broken, the two ammonia  molecules can remain in the vicinity of 
each other for some time before either the bond is reformed or the molecules diffuse 
away from each other. 
We define $N_{HB}(t)$ as the time dependent probability that
a hydrogen bond is broken at time zero but the two molecules remain in the vicinity of 
each other i.e. as nearest neighbors but not hydrogen bonded at time $t$.. 
Following previous work\cite{LC}, we can then write a simple rate equation for the 
`reactive flux' $-dC_{HB}/dt$ in terms of $C_{HB}(t)$ and $N_{HB}(t)$   
\begin{equation}
-\frac{dC_{HB}(t)}{dt}=k\,C_{HB}(t) - k'\,N_{HB}(t) \ ,
\end{equation}
where $k$ and $k'$ are the forward and backward rate constants for hydrogen
bond breaking. The inverse of $k$ can be interpreted as the average lifetime of
a hydrogen bond.

The derivative of the intermittent hydrogen bond correlation or the `reactive flux' of 
Eq.(14) is computed from the simulation results of $C_{HB}(t)$ that are
presented in Fig.~\ref{fig7}b. The probability function $N_{HB}(t)$ is also calculated
from the simulation trajectories through the following correlation function
approach~\cite{XB}
\begin{equation}
N_{HB}(t)=<h(0)[1-h(t)]h'(t)>/<h> \ ,
\end{equation}
where $h'(t)$ is unity if the N--N distance of the pair of ammonia molecules is less than
R$_{c}^{(NN)}$ at time $t$ and it is zero otherwise. The results of $N_{HB}(t)$ are shown 
in Fig..~\ref{fig7}c.
We used a least-squares fit of Eq.(14) to the simulation results of the reactive
flux, $C_{HB}(t)$ and $N_{HB}(t)$ to produce the forward and backward rate constants.
We
performed the fitting in the short time region $0<t<0.15$~ps to obtain the rate 
constants and the corresponding average hydrogen bond lifetime for the librational
relaxation and we also
carried out the fitting on the longer time region 0.25~ps $<$ t
$<$ 3~ps to calculate these quantities for slower rotational and translational
diffusional relaxation. The inverses of the corresponding forward rate constants,
which correspond to the average hydrogen bond lifetimes and which we denote as $1/k_{\rm short}$ 
and $1/k_{\rm long}$, are included in Table VII. We note that the value of 
$1/k_{\rm short}$ is very close to average hydrogen bond lifetime $\tau_{HB}$ obtained from the
continuous hydrogen bond correlation function $S_{HB}(t)$ which is not unexpected because 
both $S_{HB}(t)$
and the short-time part of the `reactive flux' captures the hydrogen bond `breaking' dynamics 
due to fast librational motion. The hydrogen bond lifetime as given by $1/k_{long}$ is 
significantly longer because it excludes the fast librational relaxation.

\section{Summary and Conclusions}

\begin{figure}
\vspace{3cm}
\begin{tabular} {c c}
 \includegraphics[width=3.5cm,angle=90]{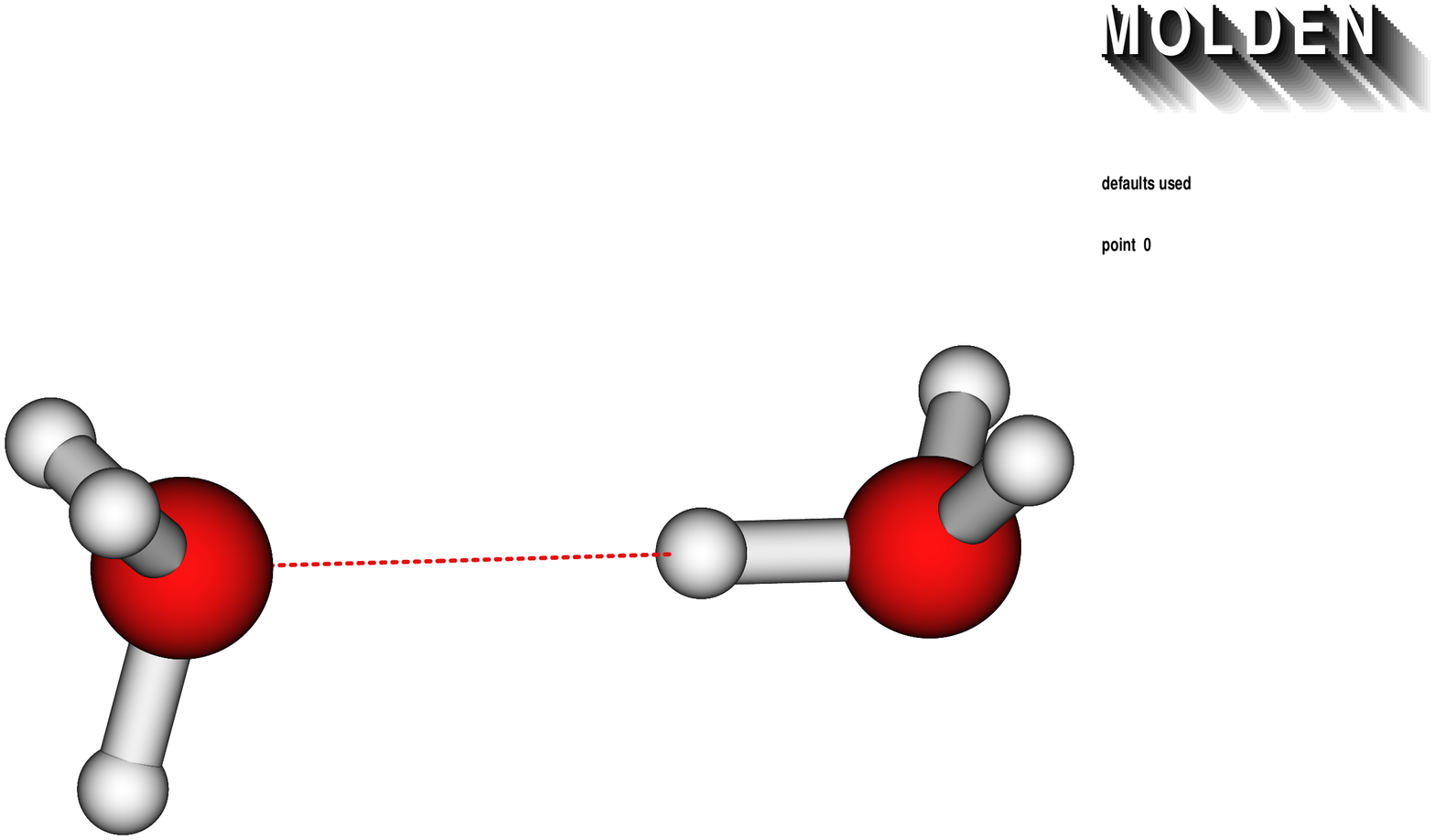} &
 \includegraphics[width=3.5cm,angle=90]{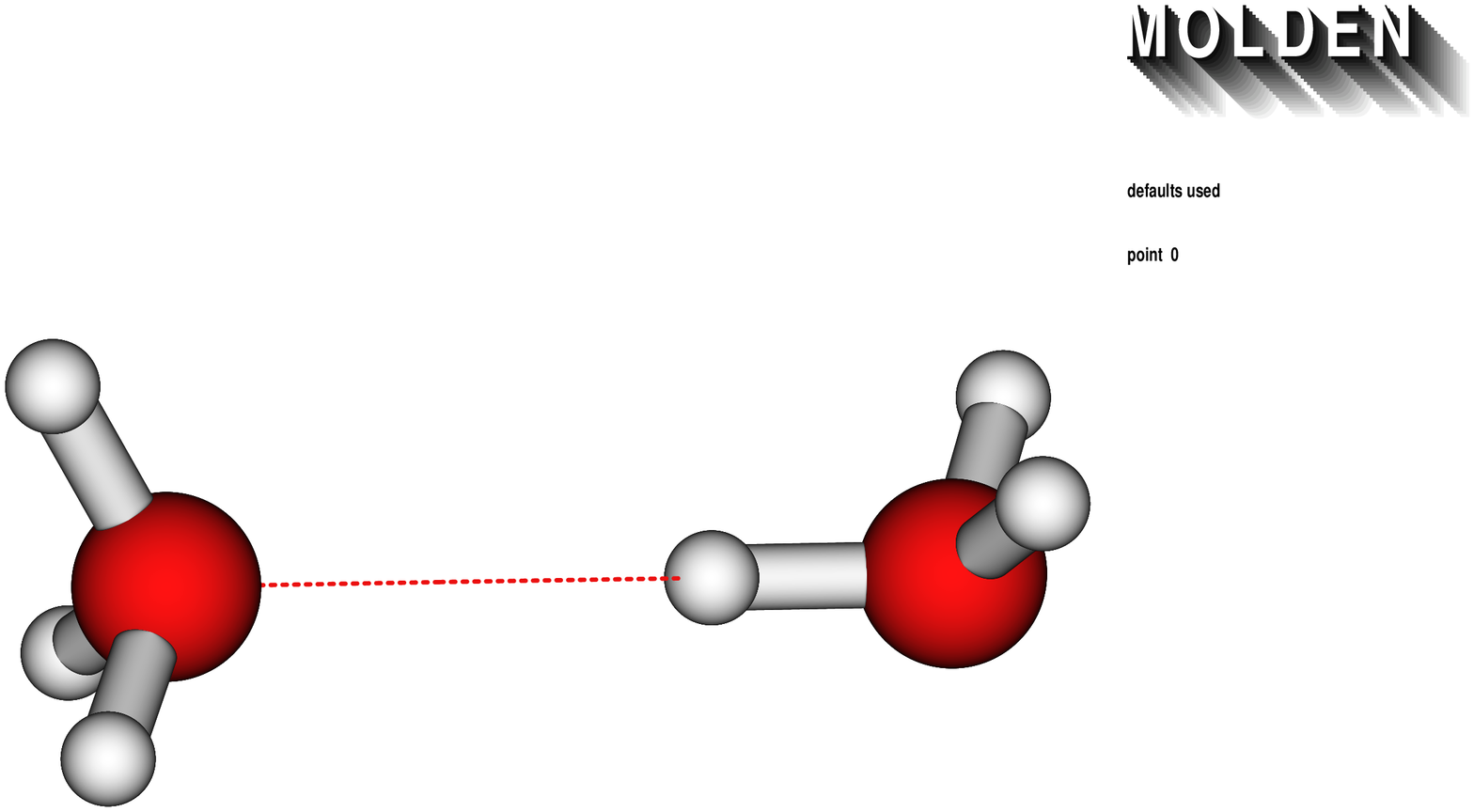} \\
 linear eclipsed &
 linear staggered \\
 Fig. 1a &
 Fig. 1b \\
 \includegraphics[width=3.5cm,angle=90]{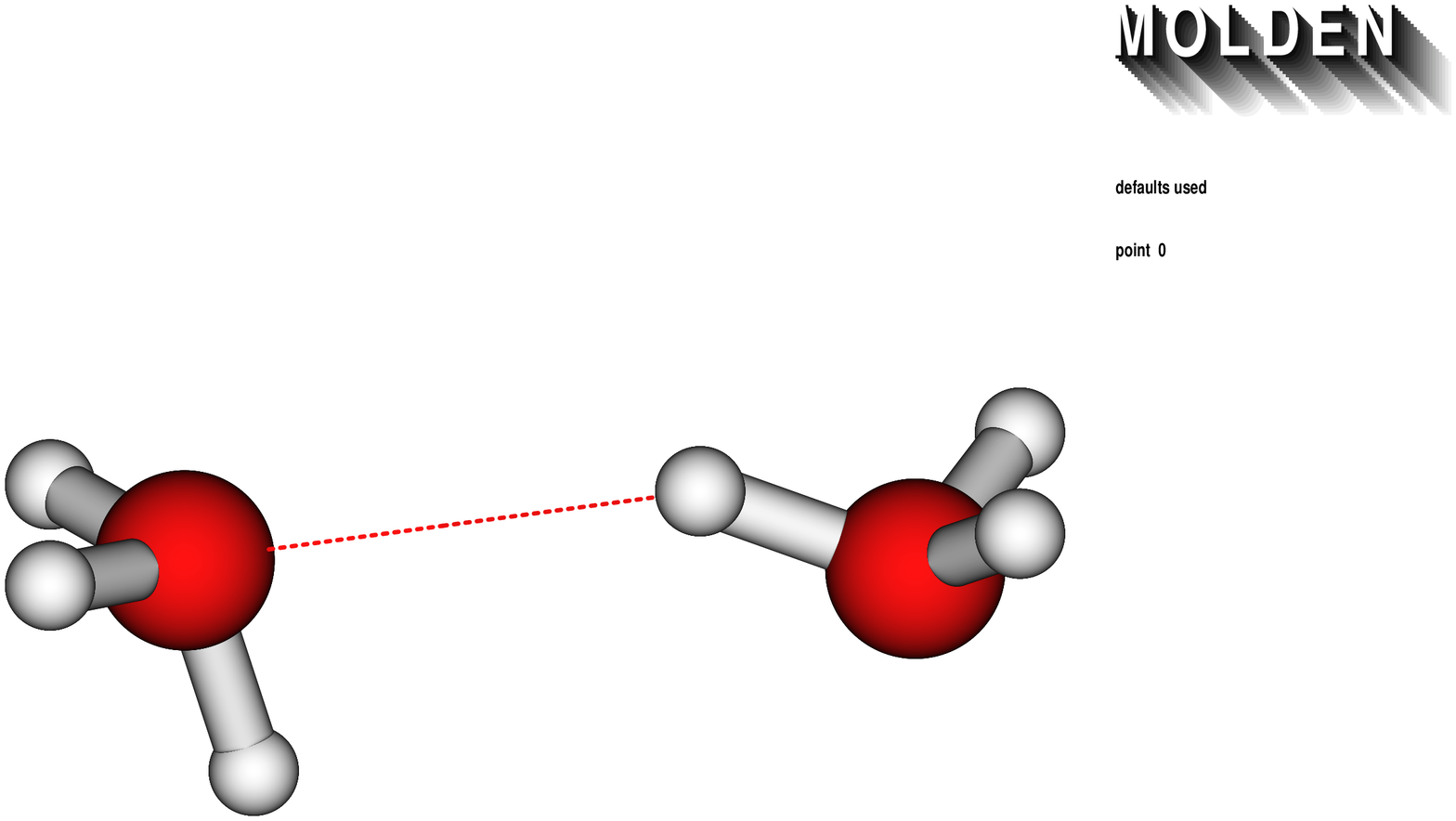} &
 \includegraphics[width=3.5cm,angle=90]{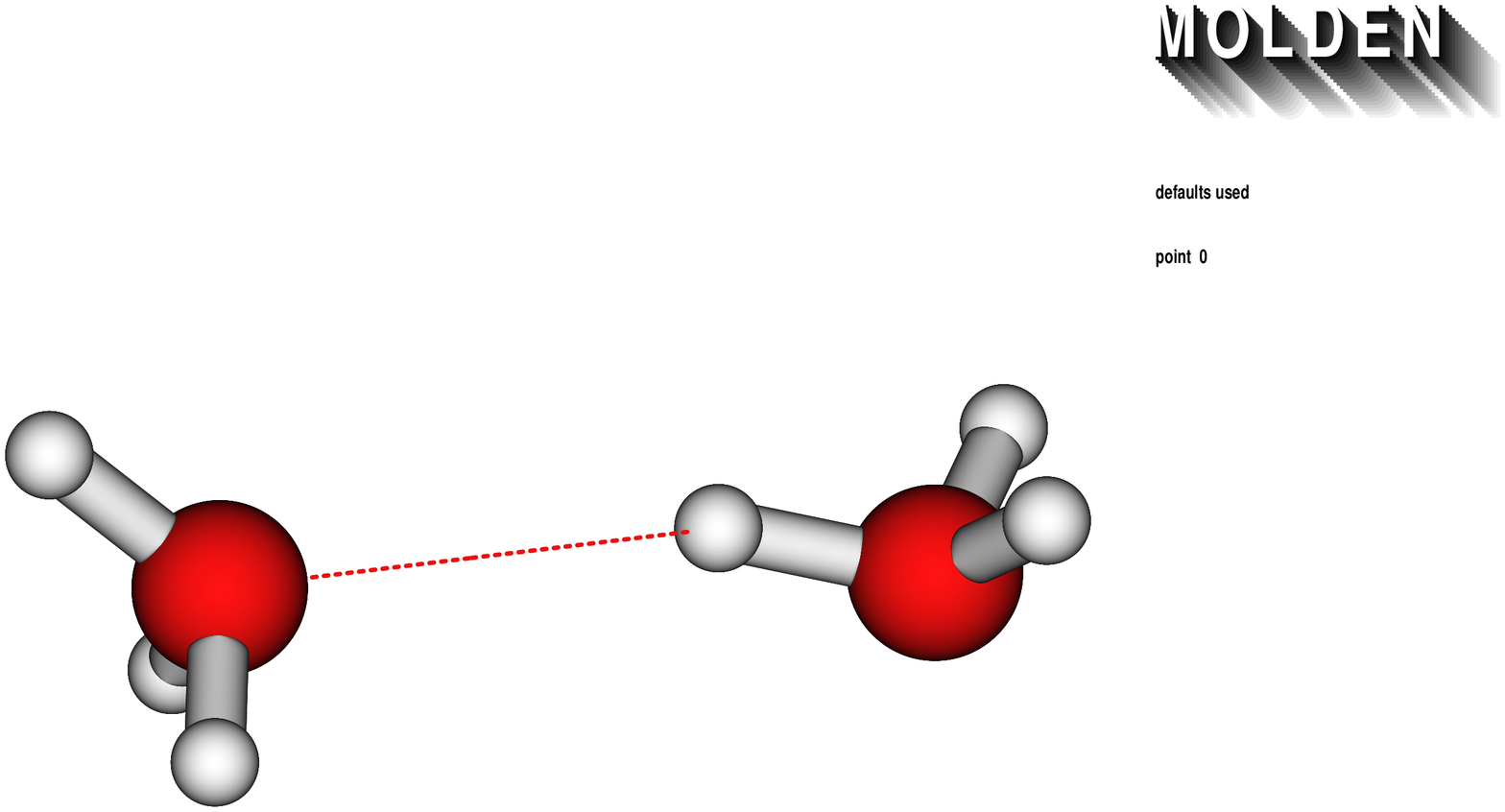} \\
 C$_s$ eclipsed &
 C$_s$ staggered \\
 Fig. 1c &
 Fig. 1d \\
 \includegraphics[width=3.5cm,angle=90]{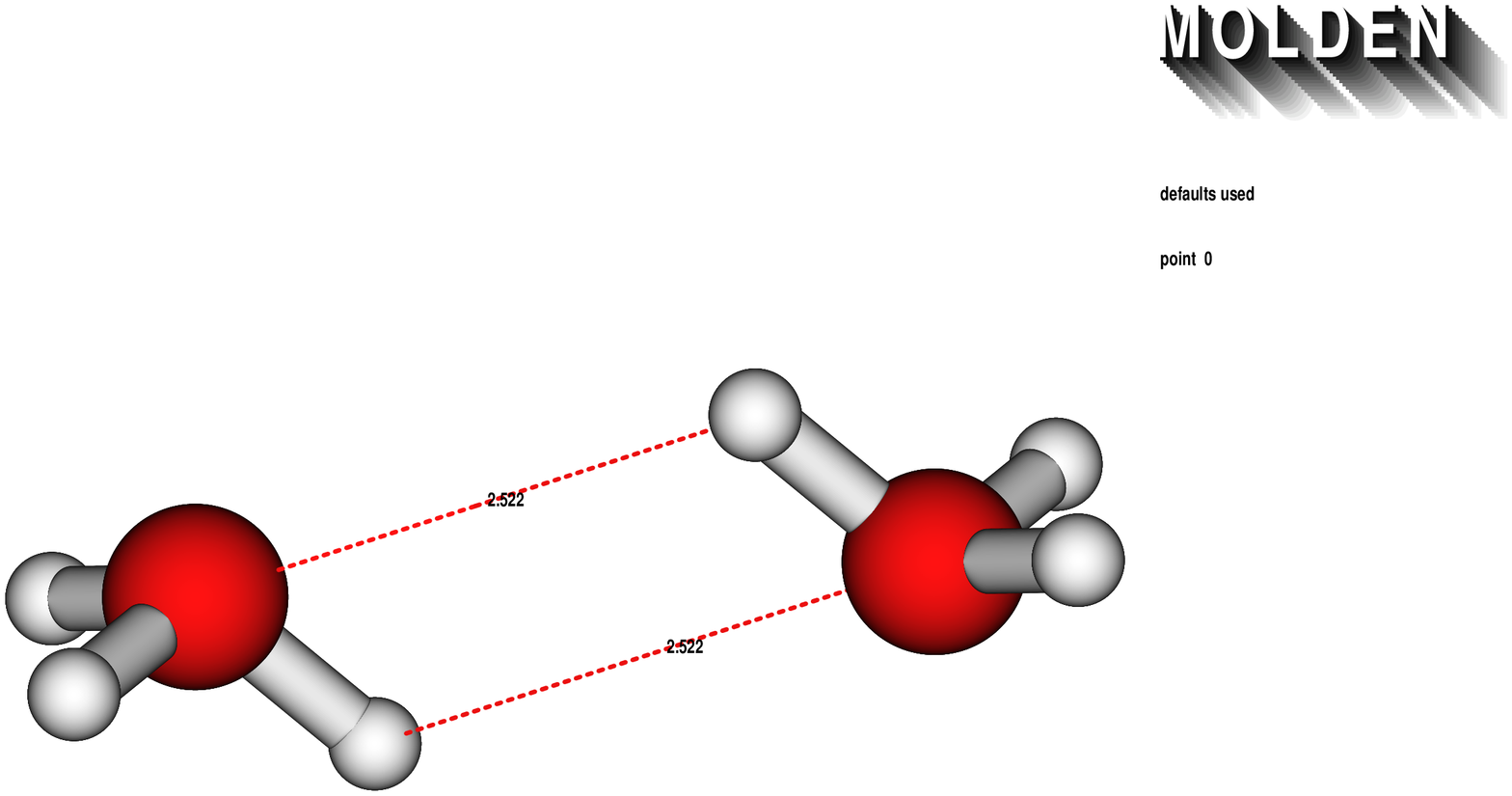} & \\
 C$_{2h}$ & \\
 Fig. 1e & \\
\end{tabular}
\caption{\label{fig1}Boese et al, Journal of Chemical Physics}
\end{figure}

\begin{turnpage}
\begin{figure}
\begin{tabular} {c c c c c}
\includegraphics[width=1.2cm,angle=90]{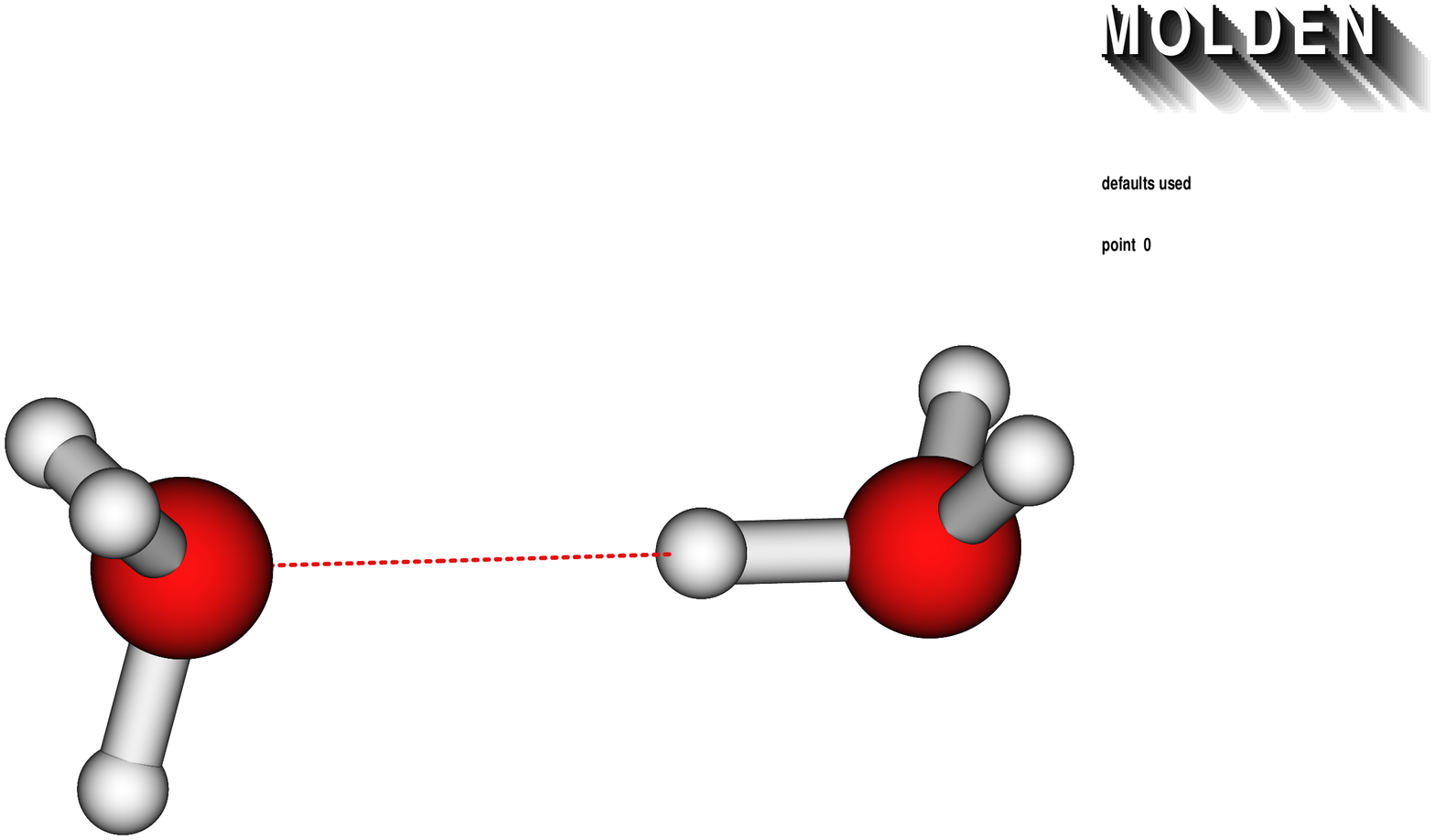} & &
\includegraphics[width=1.2cm,angle=90]{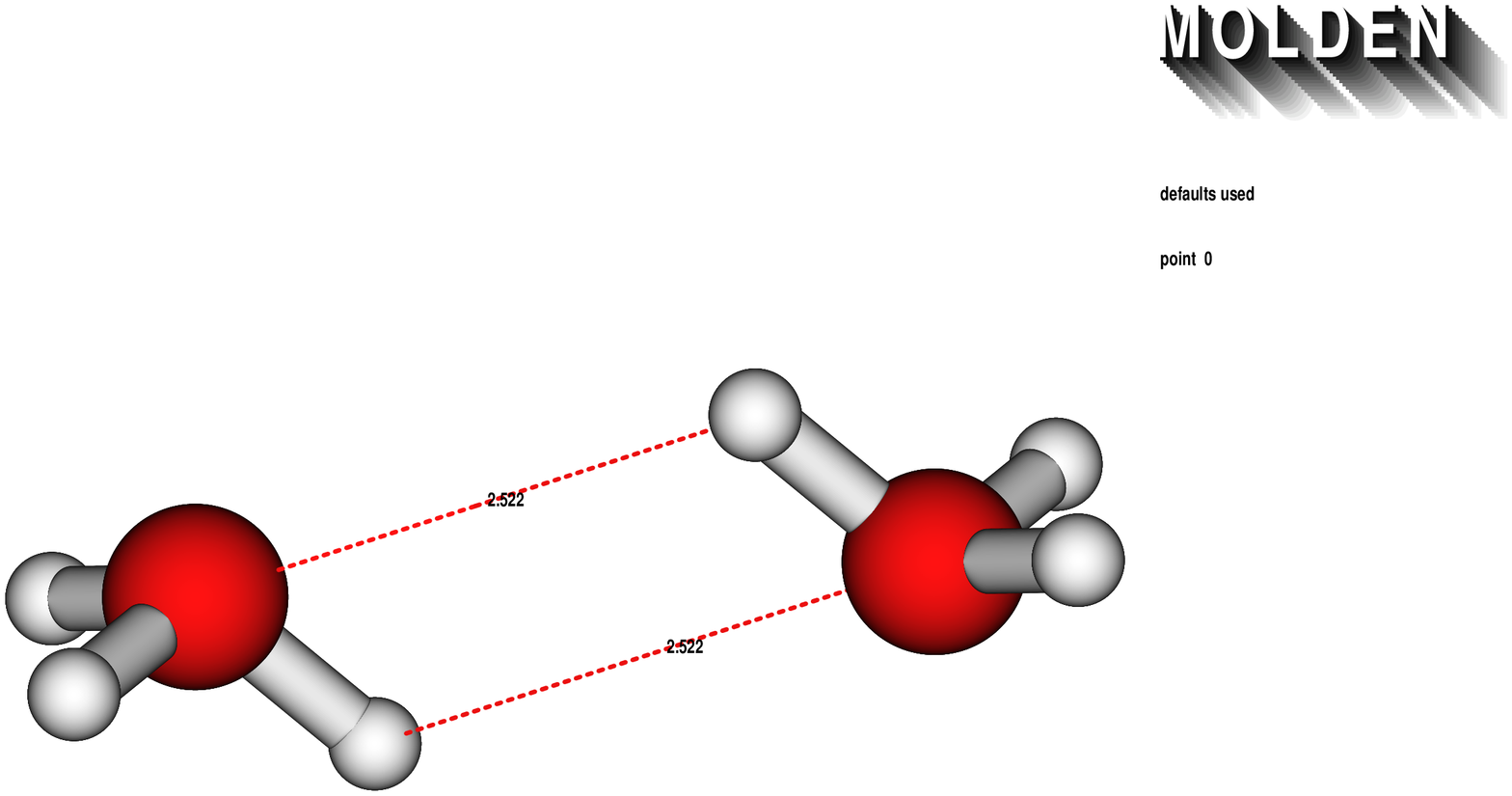} & &
\includegraphics[width=1.2cm,angle=90]{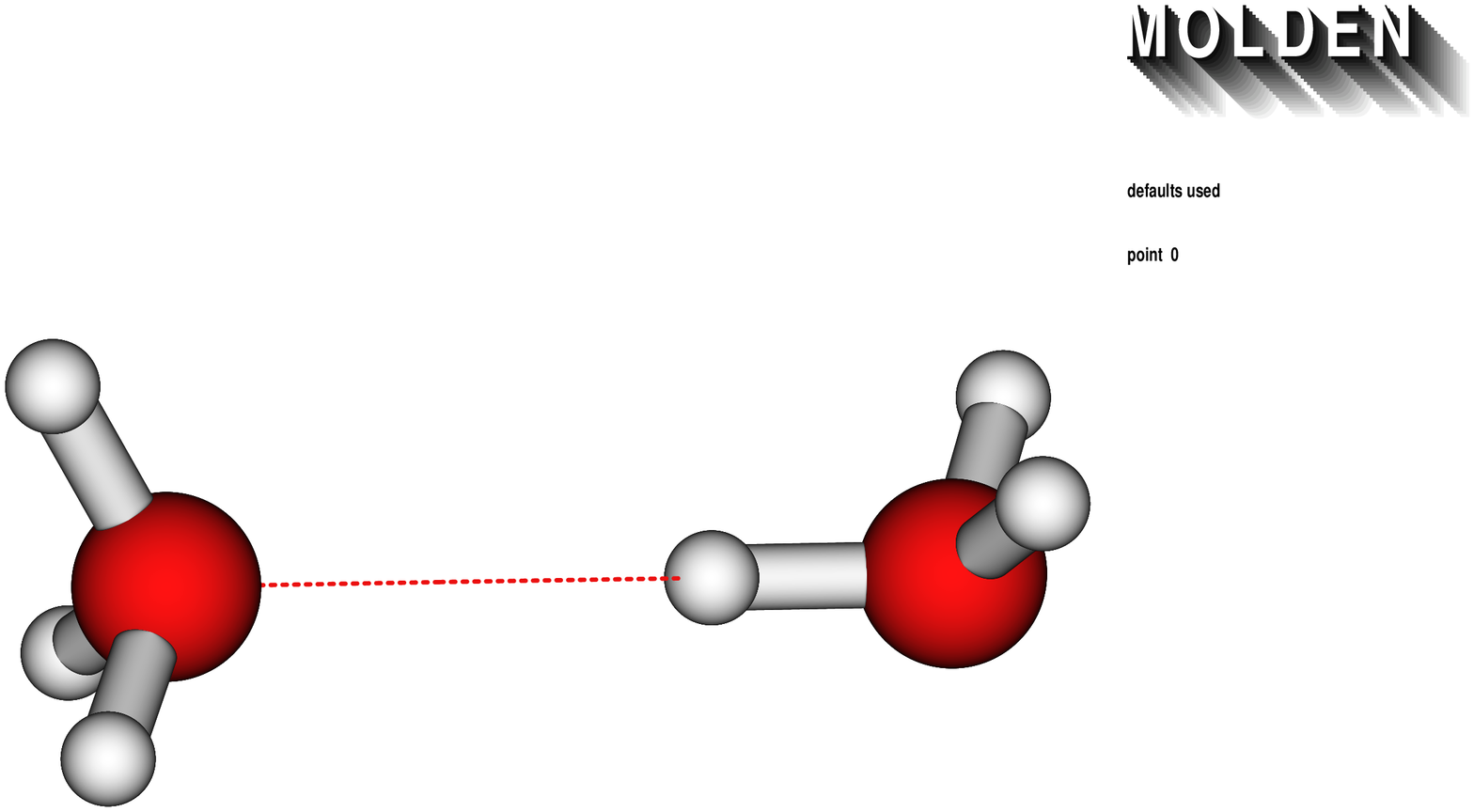} \\
\multicolumn{5}{c}{\vspace{0.8cm}}\\
\multicolumn{5}{c}{\includegraphics[width=15cm]{305335JCP.2.eps}}\\
& \includegraphics[width=1.2cm,angle=90]{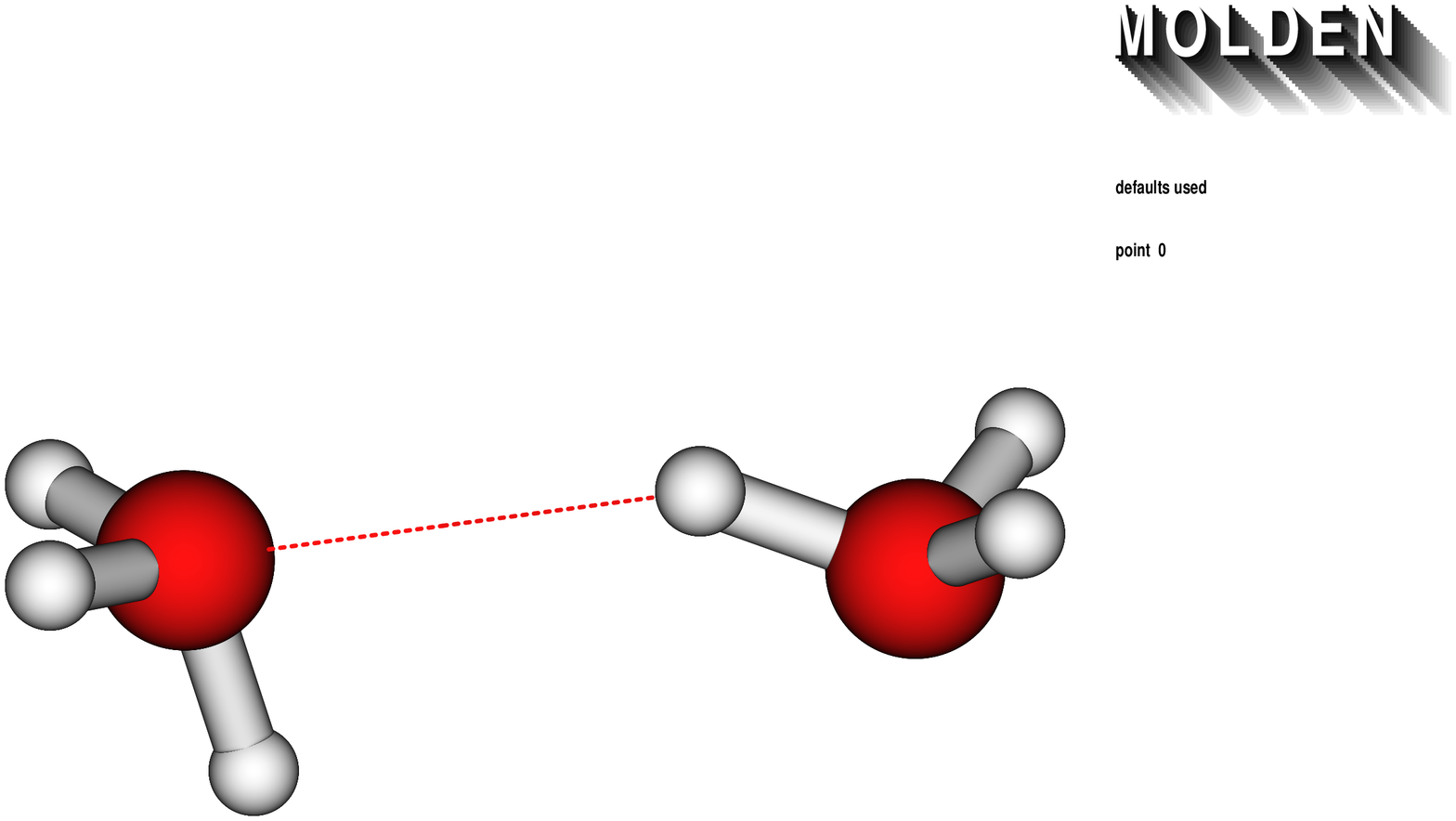} &
& \includegraphics[width=1.2cm,angle=90]{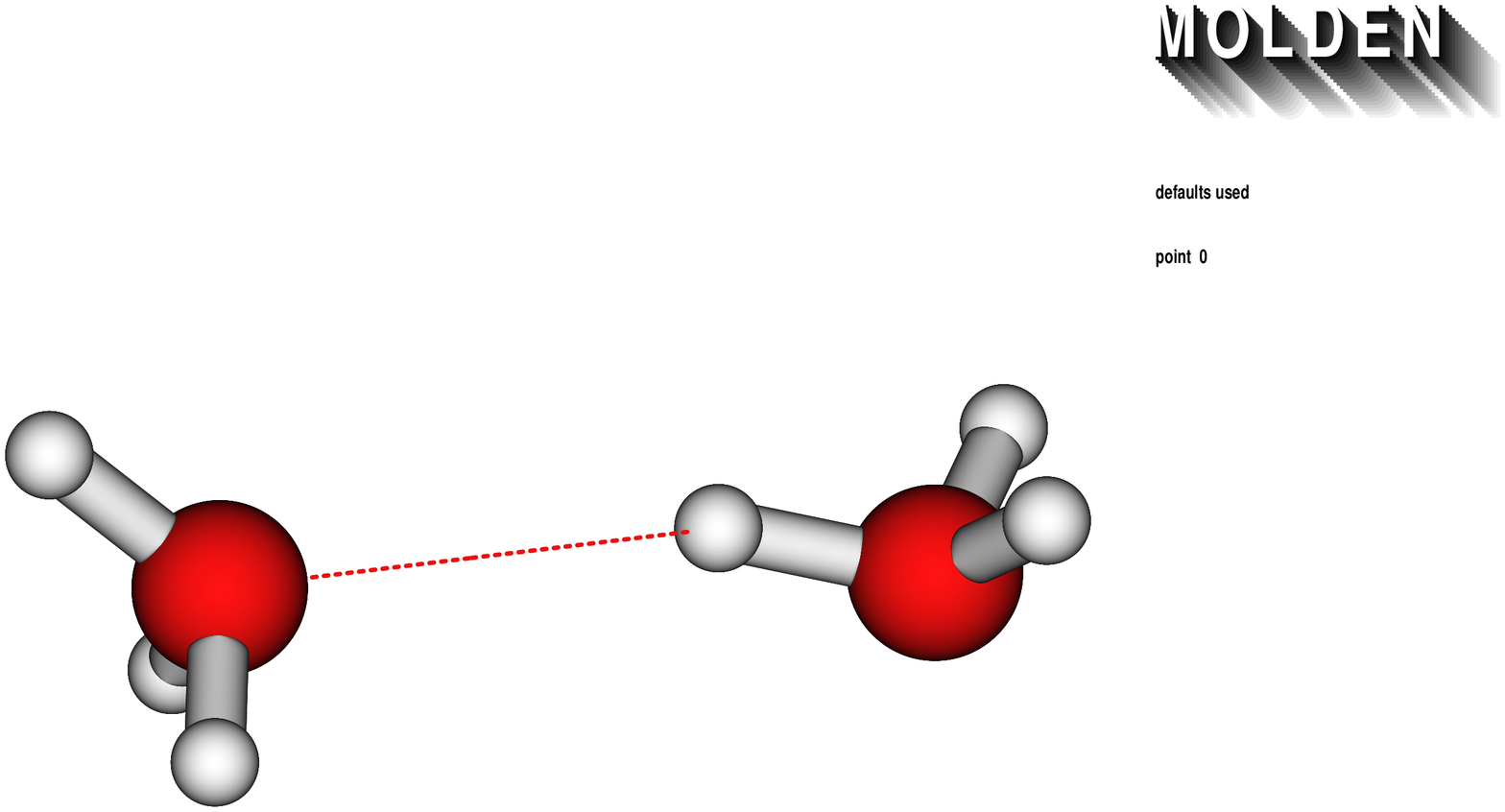} & \\
\end{tabular}
\caption{\label{fig2}Boese et al, Journal of Chemical Physics}
\end{figure}
\end{turnpage}

We have investigated the properties of ammonia from the gas phase
dimer
to the liquid state by means of 
wavefunction based quantum chemistry techniques, 
density functional theory, and {\it ab initio} molecular dynamics. 
The ammonia dimer is a particularly interesting hydrogen--bonded system,
as it is known to feature a strongly bent hydrogen bond.
In order to investigate possible consequences of this nonlinear hydrogen bond,
important points on the potential energy surface were first investigated
by means of {\it ab initio} methods up to W2--extrapolated CCSD(T) theory,
see Fig.~\ref{fig1} for the corresponding structures.
The structure and energetics of important isomers are obtained
up to unprecedented accuracy without resorting to experiment;
see Fig.~\ref{fig1} and Table~\ref{tab4} for structures, 
and Fig.~\ref{fig2} and Table~\ref{tab3} for energies.   
It is confirmed that the structure of the global minimum 
(having eclipsed $C_s$ symmetry) possesses a substantially nonlinear hydrogen bond 
with an H--N$\cdots$N angle that is predicted to be 20.7$^\circ$;
note that the corresponding angle is predicted to be 5.5$^\circ$ in the water dimer.  
The energy difference between the global minimum and the C$_{2h}$
transition state (which is the famous ``cyclic structure'') is about 3.5~cm$^{-1}$,
and the energy penalty to make the hydrogen bond artificially linear
amounts to only 70~cm$^{-1}$, i.e. 0.2~kcal/mol or 0.009~eV.
%
This implies that the potential energy surface is {\em very} flat.

However, 
when investigating the ammonia dimer using a wide variety
of available density functionals
(including important representatives of
the GGA, meta--GGA and hybrid functional families)
it was found that {\em none} of the functionals checked by us
describes the bent hydrogen bond in (NH$_3$)$_2$ satisfactorily.  
Typically, the hydrogen bond angle is too small by a factor of two
and the energy of the cyclic $C_{2h}$ structure w.r.t. the eclipsed
$C_s$ global minimum is overestimated by a factor of 10 to 30. 
Since ammonia is an important hydrogen bonded liquid, 
a density functional was developed for studying
this subtle system, with particular focus on its applicability
for condensed phase simulations in the framework
of Car--Parrinello molecular dynamics.
To this end, a novel GGA--type density functional, HCTH/407+, was developed,
with special emphasis on the nonlinear hydrogen bond 
and potential energy surface of (NH$_3$)$_2$. 
This functional yields the bent hydrogen--bonded equilibrium structure 
of the ammonia dimer,
as well as the correct energetics of the low--lying 
isomers including the energy barrier to a linear hydrogen--bond.

The performance of the new functional in describing the structural and
dynamical properties of liquid ammonia was investigated by carrying out
Car--Parrinello molecular dynamics simulations of the liquid phase. 
In addition, simulations were carried out using a 
GGA--type functional that is widely used
for describing associated liquids, BLYP, and the results of the two functionals 
are compared with experiments wherever available. 
In particular, we focused on the
atom-atom radial distribution functions, on structure and dynamics of
hydrogen bonds, and on diffusion as well as orientational relaxation processes.
Overall, both functionals are found to describe 
the structural and dynamical properties of the liquid phase reasonably well.
Importantly, it is shown that the propensity to form a strongly bent hydrogen bond
-- which is characteristic for the equilibrium structure 
of the gas--phase ammonia dimer~--
is overwhelmed by steric packing effects that clearly dominate the
solvation shell structure in the liquid state. 
Similarly, 
the propensity of ammonia molecules to form bifurcated and multifurcated
hydrogen bonds in the liquid phase is found to be negligibly small.
%
%
Thus, even functionals that lead to unreasonably linear hydrogen bonds
in the limiting case of the
in vacuo ammonia dimer, such as BLYP,
yield a good description of {\em liquid} ammonia~--
albeit for the wrong reason!

\acknowledgments
We are grateful to Professor 
Martina Havenith for pointing out the pecularities
of the ammonia dimer and for stimulating discussions.
The authors would also 
like to acknowledge Professor Nicholas Handy for enlightening
discussions, his critical comments on an early draft of the manuscript,
and his help in the very beginning of the project.
A.~D. Boese is grateful for financial support by 
the EPSRC, 
the Gottlieb Daimler- und Karl Benz-Stiftung, and 
the Feinberg Graduate School. 
A. Chandra wishes to thank Alexander von 
Humboldt~--Foundation for a Research Fellowship.
The simulations were carried out
at Rechnerverbund--NRW and at {\sc Bovilab@RUB} (Bochum).
This research was also supported by 
the Lise Meitner Center for Computational Chemistry, 
the Helen and Martin~A. Kimmel Center for Molecular Design, 
DFG, and FCI.


\newpage
\clearpage
\begin{table}
\caption{Dissociation energies $D_{\rm e}$ 
(in kJ/mol) according to W$n$ theories for several 
hydrogen bonded dimers compared to 
the best available estimates from Refs.~\cite{Klopper1,Klopper2,Wurzberger}.\label{tab1}}
\begin{tabular}{|l|l|l|l|} \hline
Molecule         &  W1   & W2    & reference     \\ \hline\hline
(HF)$_2$         & 19.13 & 19.10 & 19.10 $\pm$ 0.2 \cite{Klopper1} \\ \hline
(H$_2$O)$_2$     & 20.93 & 20.84 & 21.1 $\pm$ 0.3 \cite{Klopper2} \\ \hline
(H$_2$O)(NH$_3$) & 26..94 & 26.82 & 27.4 $\pm$ 3 \cite{Wurzberger} \\ \hline
\end{tabular}
\ \\
\ \\
\ \\
\ \\
\ \\
\end{table}

\newpage
\clearpage
\begin{table}
\caption{Relative energies (in cm$^{-1}$) according to W$n$ theories 
for the ammonia dimer at the respective CCSD(T)/A$'$PVQZ optimised structures compared
to the global minimum;
for the latter the dissociation energies $D_{\rm e}$ (in kJ/mol) into isolated optimized monomers
is reported in the last line. 
In column 2, and column 4, standard W1 and W2 results are reported, while in column 3,
a counterpoise-correction is employed on each step in the W2 calculation.
\label{tab2}
}
\begin{tabular}{|l|l|l|l|l|} \hline
Points on PES      & W1 (w BSSE) & W2(w/o BSSE) & W2(w BSSE) & best estimate \\ \hline\hline
C$_s$ (staggered)  & 23.4        &              & 23.5       & 23 $\pm$ 5 \\ \hline
C$_{2h}$           & 0.7         & 5.9          & 2.8        & 3.5 $\pm$ 3 \\ \hline
linear (eclipsed)  & 69.6        &              & 67.5       & 67 $\pm$ 5 \\ \hline
linear (staggered) & 71.8        &              & 69.9       & 70 $\pm$ 5 \\ \hline
\multicolumn{5}{|l|}{Global Minimum}   \\ \hline
C$_s$ (eclipsed)   & 13.21       & 13.15        & 13.13      & 13.1 $\pm$ 0.2 \\ \hline
\end{tabular}
\end{table}

\newpage
\clearpage
\begin{table}
\caption{Various first principles results
for the dissociation energies $D_{\rm e}$ (in kJ/mol) of the eclipsed $C_s$ global minimum
and relative energies (in cm$^{-1}$) of several structures of the ammonia dimer w.r..t.
the global minimum.
All data, except that for W2, were obtained using
an A$'$PVTZ basis set at fully optimised geometries including the counterpoise correction.
W1 and our best estimates for these values are reported in Table \ref{tab2}.
\label{tab3}
}
\begin{tabular}{|l|l|l|l|l|l|} \hline
Geometry & C$_s$ (ecl.) & C$_s$ (sta.) & C$_{2h}$     & lin (ecl.)   & lin (sta.) \\ 
Type     & De           & difference   & difference   & difference   & difference \\ \hline
Method   & in kJ/mol    & in cm$^{-1}$ & in cm$^{-1}$ & in cm$^{-1}$ & in cm$^{-1}$ \\ \hline\hline 
W2       & 13.13        & 23.5         & 2.8          & 67.5         & 69.9       \\ \hline
CCSD(T)  & 12.22        & 18.2         & 2.8          & 62.8         & 63.8       \\ \hline
CCSD     & 11.31        & 18.4         & 3.3          & 60.7         & 59.4       \\ \hline
MP2      & 12.27        & 20.3         & 2.9          & 71.8         & 72.7       \\ \hline
HF       & 7.58         & 12.6         & 23.6         & 32.3         & 34.4       \\ \hline
BLYP     & 9.11         & 5.7          & 97.8         & 36.5         & 36.4       \\ \hline
PBE      & 13.05        & 9.2          & 91..5         & 42.1         & 41.4       \\ \hline
HCTH/120 & 9.48         & 6.2          & 75.1         & 31.1         & 32.4       \\ \hline
HCTH/407 & 11.23        & 11.0         & 27.0         & 39.0         & 41.1       \\ \hline
$\tau$-HCTH & 10.96     & 5.2          & 141.8        & 29.9         & 29.6       \\ \hline
B3LYP    & 10.18        & 3.3          & 76.6         & 40.2         & 40.1       \\ \hline
B97-1    & 12.73        & 11.2         & 61.1         & 41.5         & 41.6       \\ \hline
$\tau$-HCTH hyb. & 11.27& 9.4          & 99.9         & 37.7         & 37.2       \\ \hline
HCTH/407+ & 13.18       & 18.0         &  4.0         & 51.5         & 53.6         \\ \hline
\end{tabular}
\end{table}

\newpage
\clearpage
\begin{table}
\caption{%
Various first principles results
for representative structural parameters of the ammonia dimer,
see Fig.~\protect\ref{fig1}, 
using the A$'$PVTZ basis set.\label{tab4}}
%
\begin{tabular}{|l|l|l|l|l|} \hline
Geometry & C$_s$ (ecl.)    & C$_{2h}$      & C$_s$ (ecl.) & C$_s$ (sta.) \\ 
Type     & N$\cdots$H Distance   & N$\cdots$H Distance & HNN Angle    & HNN Angle    \\ \hline
best estimate& 2.31        & 2.52          & 20.7         & 13.2         \\ \hline
CCSD(T)$^1$ & 2.302        & 2.522         & 19.86        & 13.09        \\ \hline
CCSD(T)  & 2.294           & 2.527         & 16.40        & 12.43        \\ \hline
CCSD     & 2.331           & 2.562         & 16.76        & 12.39        \\ \hline
MP2      & 2.286           & 2..520         & 17.32        & 13.28        \\ \hline
HF       & 2.541           & 2..772         & 13.41        &  9.69        \\ \hline
BLYP     & 2.341           & 2.610         &  9.73        &  8.55        \\ \hline
PBE      & 2.249           & 2.512         &  9.90        &  8.47        \\ \hline
HCTH/120 & 2.427           & 2.736         & 10.27        &  8.65        \\ \hline
HCTH/407 & 2.493           & 2.773         & 13.71        & 10.25        \\ \hline
$\tau$-HCTH & 2.311        & 2.645         &  8.40        &  7.45        \\ \hline
B3LYP    & 2.322           & 2.573         & 10.50        &  8.89        \\ \hline
B97-1    & 2.298           & 2.555         & 10.90        &  9.00        \\ \hline
$\tau$-HCTH hyb. & 2.269   & 2.549         &  9.57        &  8.24        \\ \hline
HCTH/407+& 2.493           & 2.754         & 16.95        & 11.33        \\ \hline
\end{tabular}
\end{table}
\smallskip
\hspace{1.4cm} $^1$ A$'$PVQZ basis set\\

\newpage
\clearpage
\begin{table}
\caption{Expansion coefficients of the HCTH/407+ functional compared to the
coefficients of the HCTH/407 functional.\label{tab5}}
\begin{tabular}{@{}|l|l|l|@{}} \hline 
Functional                  & HCTH/407+ & HCTH/407 \\ \hline\hline
$c_1=c_{X\sigma,0}$          &  1.08018 &  1.08184 \\ \hline
$c_2=c_{C\sigma\sigma,0}$    &  0.80302 &  1..18777 \\ \hline
$c_3=c_{C\alpha\beta,0}$     &  0.73604 &  0.58908 \\ \hline
$c_4=c_{X\sigma,1}$          & -0.4117  & -0..5183  \\ \hline
$c_5=c_{C\sigma\sigma,1}$    & -1.0479  & -2.4029 \\ \hline
$c_6=c_{C\alpha\beta,1}$     &  3.0270  &  4.4237 \\ \hline
$c_7=c_{X\sigma,2}$          &  2.4368  &  3.4256 \\ \hline
$c_{8}=c_{C\sigma\sigma,2}$  &  4.9807 &  5.6174 \\ \hline
$c_{9}=c_{C\alpha\beta,2}$   & -10.075 & -19.222 \\ \hline
$c_{10}=c_{X\sigma,3}$       &  1.3890 & -2.6290 \\ \hline
$c_{11}=c_{C\sigma\sigma,3}$ & -12.890 & -9.1792 \\ \hline
$c_{12}=c_{C\alpha\beta,3}$  &  20.611 &  42.572 \\ \hline
$c_{13}=c_{X\sigma,4}$       & -1.3529 &  2.2886 \\ \hline
$c_{14}=c_{C\sigma\sigma,4}$ &  9.6446 &  6.2480 \\ \hline
$c_{15}=c_{C\alpha\beta,4}$  & -29.418 & -42.005 \\ \hline
\end{tabular}
\end{table}

\newpage
\clearpage
\begin{turnpage}
\begin{table}
\caption{Comparison of the errors of the 407+ fit set for the HCTH/407+ and previous published functionals.
Also included are the errors for properties of nine hydrogen-bonded systems (see text)..\label{tab6}}
\begin{tabular}{|l|l|l|l|l|l|} \hline
 & \multicolumn{2}{|l|}{ 407+ Set of molecules }&\multicolumn{3}{|l|}{ Hydrogen-Bonded Systems } \\\hline
Property & RMS Energy & $\sum$ Gradient & RMS Dissociation Energy & RMS H-Bond Shift & RMS Frequency Shift \\ \hline
Unit & [kcal/mol] & [a.u.] & [\%] & [\%] & [\%] \\ \hline\hline
HCTH/120 & 9.1    & 11.51        & 8.5        & 26.0       & 29.1       \\ \hline
HCTH/407 & 7.8    & 11.91        & 7.5        & 16.7       & 14.8       \\ \hline
HCTH/407+& 8.0    & 11.96        & 10.3       & 16.8       & 12.8       \\ \hline
B3LYP    & 9.4    & 12.00        & 11.0       & 30.2       & 35.0       \\ \hline
BLYP     & 9.7    & 19.30        & 16.4       & 42.1       & 42.9       \\ \hline
BP86     & 16.4   & 17.05        & 22.3       & 81.9       & 68.3       \\ \hline
\end{tabular}
\end{table}
\end{turnpage}

\newpage
\clearpage
\begin{table}
\caption{Dynamical properties of liquid ammonia. The diffusion coefficient,
relaxation times and the inverse rate constants are expressed in units of
10$^{-5}$\,cm$^{2}$sec$^{-1}$, ps and ps, respectively. 
Note that {\em all} numbers given refer to the fully deuterated species ND$_3$. 
\label{tab7}}
\begin{tabular}{@{}|l|l|l|l|@{}} \hline 
Quantity           & HCTH/407+ & BLYP     & Experiment \\ \hline\hline
D                  &  5.23     & 5.44     & 8.75$^a$, 5.95$^b$        \\ \hline
$\tau_{1}^{dipole}$  &  0.52     & 0.85     &         \\ \hline
$\tau_{1}^{NH}$      &  0.37     & 0.58     &         \\ \hline
$\tau_{1}^{HH}$      &  0.34     & 0.54     &         \\ \hline
$\tau_{2}^{dipole}$  &  0.18     & 0.27     &         \\ \hline
$\tau_{2}^{NH}$      &  0.15     & 0.22     &         \\ \hline
$\tau_{2}^{HH}$      &  0.15     & 0.22     &         \\ \hline
$\tau_{2}^{\rm exp}$ &       &      &  0.26$^a$, 0.35$^b$        \\ \hline
$\tau_{HB}$          &  0.07     & 0.10     &         \\ \hline
$1/k_{\rm short}$      &  0.10     & 0.16     &         \\ \hline
$1/k_{\rm long}$      &  0.62     & 0.60     &         \\ \hline
\end{tabular}
\end{table}
\smallskip
\hspace{3.75cm} $^a$ at 275~K, ~$^b$ at 252~K\\

\newpage
\clearpage
\noindent
Fig. 1.  The five relevant optimized
structures of the ammonia dimer;
dotted lines are only guides to the eye..

\ \\
\noindent
Fig. 2.
The potential energy profile obtained from W2 Theory 
(accurate reference calculations) in comparison to that obtained from
HCTH/407+ and BLYP functionals 
at the five optimized structures, see insets, from Fig.~\protect\ref{fig1}.
Note that the profiles obtained from other GGA functionals
such as PBE are qualitatively
similar to that of BLYP. 

\ \\
\noindent
Fig.3.
The atom-atom radial distribution functions. The solid, dashed and the dotted
curves show the HCTH/407+, BLYP, and the experimental~\cite{RNR,Soper} results, respectively.

\ \\
\noindent
Fig.4.
The fraction $P(n)$ of molecules having a certain number of 
accepted $n_{\rm acceptor}$ and donated  $n_{\rm donor}$ 
hydrogen bonds, see panels (a) and (b), respectively. 
Panel (c) shows the fraction of hydrogen atoms that donates a certain number
of hydrogen bonds.
The squares and the triangles represent the results
for the HCTH/407+ and BLYP functionals, respectively, and the dashed lines are drawn to
guide the eye.

\ \\
\noindent
Fig.5.
The distribution of the cosine of the N$\cdots$N--H angle for H atoms that belong to
the nearest neighbours. Panel (a) shows the distribution for hydrogen bonded H atoms 
(i.e.. R$^{(NH)}$ $<$ 2.7~\AA) and panels (b) and (c) show, respectively, the results
for non--hydrogen--bonded H atoms that appear in the region 2.7~\AA~ $<$ R$^{(NH)}$ $<$ 3.7~\AA~
and 3.7~\AA~ $<$ R$^{(NH)}$ $<$ 4.7~\AA.
The solid and the dashed curves show the results for the HCTH/407+ and BLYP
functionals, respectively.

%
%

\ \\
\noindent
Fig.6.
The time dependence 
various orientational correlation functions,
see text for definitions and further details. 
The solid and the dashed curves show the results for the HCTH/407+ and BLYP
functionals, respectively.


\ \\
\noindent
Fig.7.
The time dependence of various hydrogen bond correlation functions, 
see text for definitions and further details. 
The different curves are labelled as in Fig.6.

\newpage
\pagestyle{empty}
\clearpage

\newpage
\pagestyle{empty}
\clearpage

\newpage
\pagestyle{empty}
\clearpage
\begin{figure}
\includegraphics{305335JCP.3.eps}
\caption{\label{fig3}Boese et al, Journal of Chemical Physics}
\end{figure}

\newpage
\pagestyle{empty}
\clearpage
\begin{figure}
\includegraphics{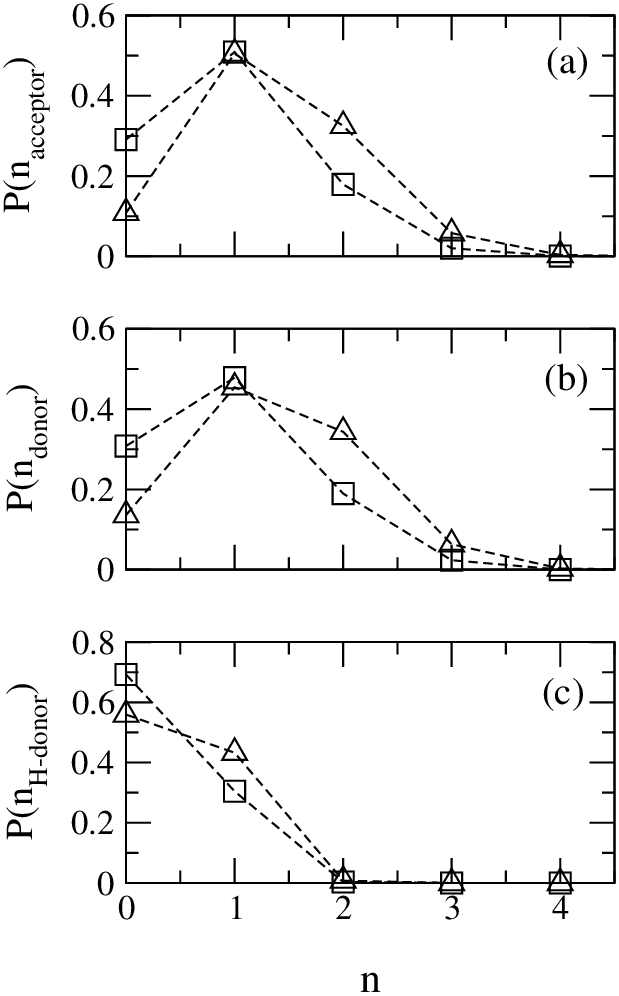}
\caption{\label{fig4}Boese et al, Journal of Chemical Physics}
\end{figure}

\newpage
\pagestyle{empty}
\clearpage
\begin{figure}
\includegraphics{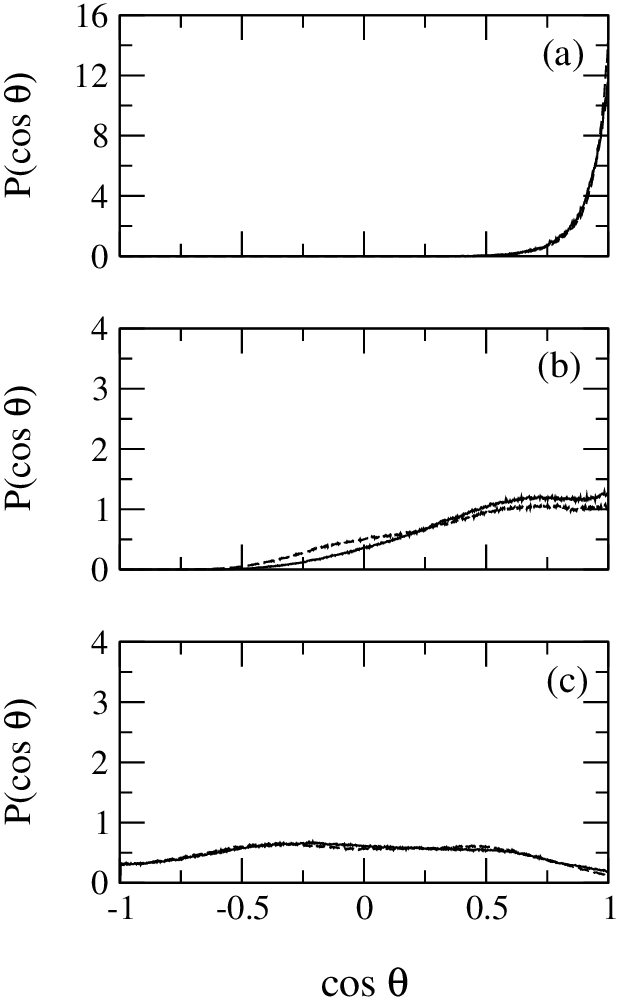}
\caption{\label{fig5}Boese et al, Journal of Chemical Physics}
\end{figure}

\newpage
\pagestyle{empty}
\clearpage
\begin{figure}
\includegraphics{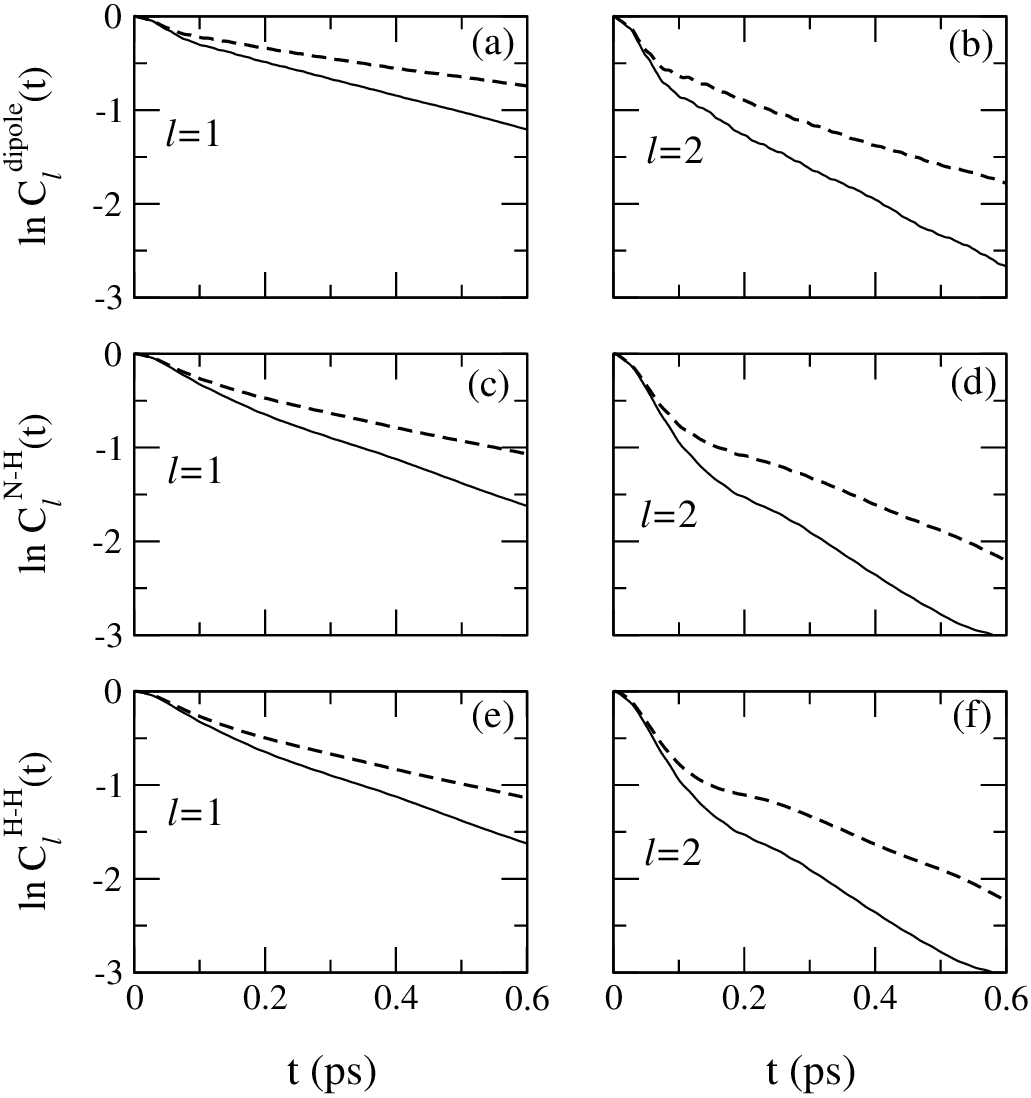}
\caption{\label{fig6}Boese et al, Journal of Chemical Physics}
\end{figure}

\newpage
\pagestyle{empty}
\clearpage
\begin{figure}
\includegraphics{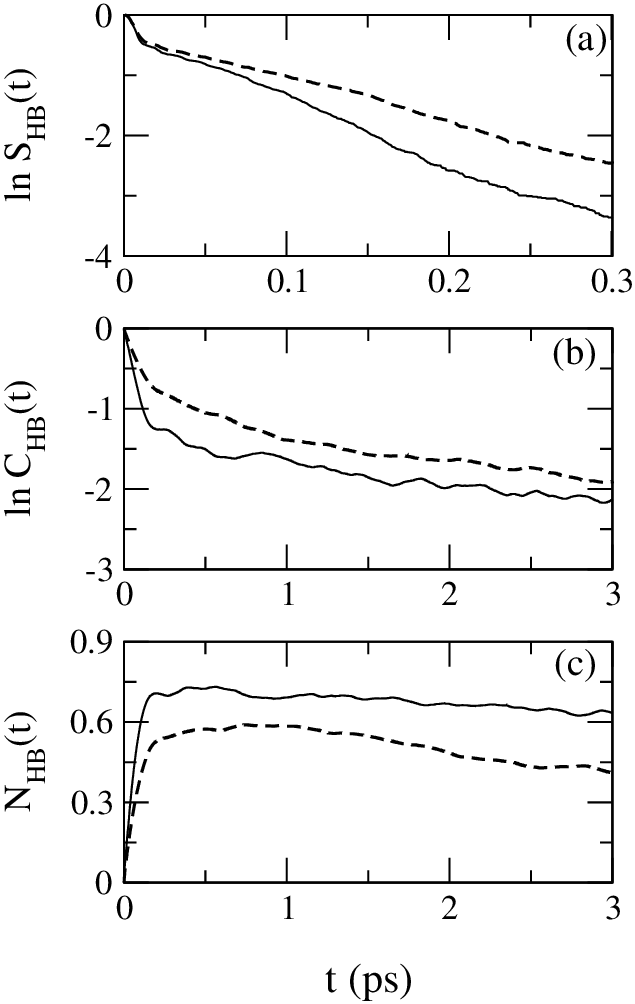}
\caption{\label{fig7}Boese et al, Journal of Chemical Physics}
\end{figure}


\begin{thebibliography}{200}
\bibitem{noteHB} Note that we adopt here Steiner's terminology~\protect\cite{steiner02}:
	(i) in a hydrogen bond ${\cal D}$--H$\cdots$${\cal A}$ 
	the group ${\cal D}$--H is called the 
	(proton) donor and ${\cal A}$ is called the (proton) acceptor;
	(ii) the interaction of one donor with two or more acceptors 
	${\cal A}'$, ${\cal A}''$, $\dots$~
	leads to so--called bifurcated 
	\mbox{
	\setlength{\unitlength}{1pt}
	\begin{picture}(56,20)
	\put(0,0){$\mathcal{D}$}
	\put(11,4){\line(1,0){6}}
	\put(18,0){H}
	\multiput(28,8)(6,2){3}{.}
	\put(44,10){$\mathcal{A}'$}
	\multiput(28,0)(6,-2){3}{.}
	\put(44,-10){$\mathcal{A}''$}
	\end{picture}
	}
	\rule[-14pt]{0pt}{36pt}
	%
	or multifurcated hydrogen bonds.
	Note that sometime the cyclic $C_{2h}$ structure of (NH$_3$)$_2$
	according to Fig.~\protect\ref{fig1}e
	is also termed ``bifurcated'', which we do not do here.
\bibitem{Soper} The experimental RDFs that are plotted in Fig.~\ref{fig3} 
	were derived from the diffraction 
	data of Ref.~\protect\cite{RNR}
	but using a more sophisticated 
	procedure; private communiation of A..K. Soper.
\end{thebibliography}
\end{document}